\documentclass[aps,showpacs,twocolumn,toolkits]{revtex4}
\usepackage{mathrsfs}
\usepackage{amsmath, amssymb, amsfonts, bbm}
\usepackage{graphics,epsfig,color,subfigure}
\begin{document}

\title{ Baryons and baryonic matter in the large Nc and heavy quark limits}
\author{Thomas D. Cohen}

\email{cohen@physics.umd.edu}

\affiliation{Maryland Center for Fundamental Physics and the Department of Physics, University of Maryland,
College Park, MD 20742-4111}

\author{Nilay Kumar}

\email{nilay@umd.edu}

\affiliation{Montgomery Blair High School, 51 University Boulevard East,
Silver Spring, MD 20901-2451}

 \author{Kamal K. Ndousse}
 \email{ kandouss@mit.edu}
 \affiliation{
Massachusetts Institute of Technology,
77 Massachusetts Ave,
Cambridge, Massachusetts 02142  }
\date{\today}


\begin{abstract}
This paper explores properties of baryons and finite density baryonic matter in an artificial world in which   $N_c$, the number of colors, is large and the quarks of all species are degenerate and much larger than $\Lambda_{\rm QCD}$.   It has long been known that in large $N_c$  QCD, baryons composed entirely of heavy quarks are accurately described in the  mean-field approximation.   However, the detailed properties of baryons in the combined large $N_c$ and heavy quark limits have not been fully explored.   Here some basic properties of baryons are computed using a variational approach.    At leading order in both the large $N_c$ and heavy quark expansions  the baryon mass is shown to be $ M_{\rm baryon}  \approx N_c  M_Q \left (1 - 0.05426 \,  \tilde{\alpha}_s^2  \right )$  where $\tilde{\alpha}_s \equiv  N_c {\alpha}_s $.  The baryon form factor is also computed.   Baryonic matter, the analog of nuclear matter in this artificial world, should also be well described in the mean-field approximation.   In the special case where all baryons have an identical spin flavor structure,  it is shown that in the formal heavy quark and large $N_c$ limit interactions between baryons are strictly repulsive at low densities.   The energy per baryon is computed in this limit and found to be exponentially small.  It is shown that when the restriction to baryons with an identical spin-flavor structure is dropped,  a phase of baryonic matter exists with a density of $2 N_f$ times that for the restricted case but with the same energy (where $N_f$ is the number of degenerate flavors).  It is shown that this phase is at least metastable.    \end{abstract}

\pacs{11.15.Pg,12.39.Hg,14.20.-c}

\maketitle

\pagebreak

\section{Introduction}
One of the central theoretical problems in nuclear physics is the understanding of the properties of cold nuclear matter directly from QCD.  However, this problem is quite difficult since, unlike the study of QCD at non-zero temperatures, QCD at non-zero baryon chemical potential cannot be studied using standard lattice techniques due to the presence of a fermion sign problem\cite{FSP}.  Indeed, from the point of view of the functional integral of QCD it is not even understood why the system at zero temperature remains in the vacuum state for non-zero chemical potentials below the critical value for producing nuclear matter; this is the so-called Silver Blaze problem\cite{SB}.   Given this situation it would be useful to try to gain some intuition into the problem via the study of limiting cases of QCD for which the baryon matter problem  (the generalization of the nuclear matter problem)  becomes tractable.

One obvious   approach for seeking tractable limits, is to start with the large $N_c$ limit of QCD.  It has long been recognized that certain aspects of QCD greatly simplify in the large $N_c$ limit\cite{'tHooft:1973jz}.   At the fundamental level of quarks and gluons, the large $N_c$ theory is dominated by planar graphs, while at the phenomenological level, the zero-baryon sector of the theory becomes a weakly coupled theory of mesons and glueballs.   Baryons are more complicated.  One way to proceed is  to  consider a theory in which one flavor of quark is in the two-color antisymmetric representation and other  flavors are in the color fundamental representations\cite{CR,ASV}.  Such a theory has the virtue that 3-quark baryons exist at any $N_c$.  However, such an approach badly breaks flavor symmetry and thus is an unnatural generalization of the $N_c=3$ world\cite{CCL}.   An alternative approach, pioneered by Witten\cite{Witten:1979kh}, is to respect flavor symmetry and consider baryons which contain $N_c$ quarks and deal with the combinatoric complications that arise due to this.  We will follow the `t Hooft-Witten large $N_c$ limit here in which gluons are in the adjoint representation and quarks are in the fundamental.

Unfortunately, despite its simplifications large  $N_c$ QCD is not tractable in general.  Thus,   it cannot be used directly to gain insight about baryonic matter.  To proceed further, additional simplifications are needed.  One  approach is to restrict attention to large $N_c$ QCD in 1+1 dimensions---the  `t Hooft model\cite{Hooft2}.   Meson spectroscopy in this model is tractable: it can be obtained from solutions to a simple integral equation which can be evaluated numerically.  Underlying this simplification is the fact that the gluon degrees of freedom are non-dynamical in 1+1 degrees; there are electric components of the gluon field strength but by construction no magnetic ones and thus gluons cannot propagate.  Recently, there have been significant advances in the  understanding of baryon matter in the  `t Hooft model \cite{Bringoltz,Narayanan}.

One might  similarly hope to find tractable regimes in 3+1 dimensional large $N_c$ QCD where the dynamics associated with the gluon sector is absent or highly restricted.  This  occurs if the characteristic momentum flowing through gluons is much larger than $\Lambda_{\rm QCD}$ .  In this case,  the running coupling is small and dynamical gluon effects are suppressed.  One way to achieve this is in  the regime of  very high density or equivalently chemical potential, $\mu$ .  In that case, $\mu$ provides a large scale which can push the typical momentum flowing through gluons well above $\Lambda_{\rm QCD}$.  Large $N_c$ analyses of the high $\mu$ regime have been done  both for the usual `t Hooft large $N_c$ limit (with quarks in the fundamental representation) \cite{SS} and for ${\rm QCD}_{\rm AS}$---a variant in which quarks are in the two index anti-symmetric representation---which represents a different extrapolation from $N_c=3$ \cite{BCC}.  For the case of the `t Hooft large $N_c$ limit it was shown that a DGR\cite{DGR} instability occurs,  indicating that the ground state cannot be of the BCS type but rather breaks translational symmetry\cite{SS}.   However, in ${\rm QCD}_{\rm AS}$ the DGR instability can be shown not to occur and it is plausible that the ground state is of the BCS type\cite{BCC}.   The differences between these two examples illustrate a critical point.  In going to extreme limits to gain tractability, one may well push the system to a regime which differs  in important qualitative ways from QCD at $N_c=3$ in less extreme conditions.  The existence of a DGR instability in the 't Hooft large $N_c$ limit  might be viewed as an artifact of the large $N_c$ limit.  
The regime of interest involves a double limit: $\mu \rightarrow  \infty$, $N_c \rightarrow \infty$.  It is not obvious that these two limits commute and, in fact, they do not: as stressed in ref.~\cite{MP} the behavior depends sensitively on whether the the large $N_c$ limit or the high density limit is taken first.

This paper focuses on another regime of large $N_c$ QCD for where dynamical gluon effects are suppressed:  the case where the quark masses (which are taken to be degenerate) are much larger than $\Lambda_{\rm QCD}$.  In this case the characteristic momentum flowing through dynamical gluons is typically $\alpha_s M_Q$.  The quarks, being very heavy, move nonrelativistically in this regime.  The gluons become nondynamical: they give rise to a color-Coulomb interaction between the quarks, and effects of  propagating gluons are suppressed.   It is thus not surprising that this regime provides a tractable setting for the study of both single baryons and baryonic matter.

The combined heavy quark and large $N_c$ limit has a long history.  Indeed, Witten's classic paper  on the properties of baryons in large $N_c$ QCD is based on an analysis of baryons in this combined limit;  it is first argued that this limit is accurately described in a mean-field approximation\cite{Witten:1979kh}.  From this, the $N_c$ scaling of various baryon properties  were deduced.  Finally, Witten argued that these properties were generic and remained valid even away from the heavy quark limit.   Witten's reasoning shows the power of using tractable regimes to learn rather deep things about large $N_c$ QCD more generally.

This paper explores baryons and cold baryonic matter in the combined large $N_c$ and heavy quark limits.  The problem of baryons is in a general sense quite well understood and the logic underlying Witten's mean-field approach to baryons is extremely well known.  However,  while the  formulation is over three decades old, to the best of our knowledge the equations have never been solved numerically.  It is important to do so if for no other reason than completeness---it remains one of the only regimes of QCD which is solvable.   There is also  some phenomenological utility in doing this---the $\Omega_{b}$ baryon (composed of three bottom quarks) might be regarded as being relatively near this  in this limit.  While the  $\Omega_{b}$ has not been observed to date, there is little doubt about its existence  in the real world\cite{GomshiNobary:2005ur}.

The second thrust of this paper is the problem of cold baryonic matter.  We treat the problem of infinite baryon  matter using the mean-field approach suggested by Witten.  The general solution to this problem has yet to be solved and it is not immediately clear whether baryonic matter in the combined limit saturates ({\it i.e.}, is self-bound).  However, we can show that  there exists a phase at low density which consists of weakly interacting nucleons.   In this regime the  nucleus-nucleus potential is due to the Pauli principle and strictly repulsive; however, the repulsion is exponentially small.  We can compute the equation of state for cold matter in this phase.   It is unknown whether this phase is absolutely stable.  It is known, however, to be at least  metastable with a very long lifetime.  The question of absolute stability will need to be investigated in future work.

 \section{The baryon in the combined large $N_c$ and heavy quark limits\label{1baryon}}
\subsection{The mean-field energy functional \label{bosonic} }

The general large $N_c$ scaling rules for baryons were originally deduced by Witten \cite{Witten:1979kh}. The analysis  in ref.~\cite{Witten:1979kh} begins by considering the special case where the quarks constituting the baryon are heavy. The formalism for the case of heavy quarks is rigorous.  Somewhat more heuristic arguments were then invoked to suggest that the $N_c$ scaling results for the case of heavy quarks should also hold  when the quarks are light. This allows for the determination of  the large $N_c$ scaling rules for baryon in the phenomenologically interesting regime of nearly chiral quarks.  Here we will stick to the heavy quark regime.  The analysis of this regime  in ref.~\cite{Witten:1979kh} was based largely on physical reasoning.  In this subsection a somewhat more formal derivation will be given.  Aspects of this more formal treatment help clarify the nature of the mean-field approximation for the baryon.  Moreover, it will prove extremely useful in formulating the baryonic matter calculation in the following section.

The starting point of the analysis in ref.~\cite{Witten:1979kh} is that in the heavy quark regime, quark-antiquark pairs are suppressed and the quarks behave nonrelativistically.  Color magnetic couplings are also suppressed since the color magnetic moment scales as the inverse of the  quark mass.   Moreover, gluon-gluon interaction effects are suppressed since the characteristic size scale of the system is small and thus $\alpha_s$ at this scale is small (albeit only logarithmically).   Thus for heavy quarks, a baryon is accurately described as $N_c$ quarks in a color singlet state interacting with each other through a color Coulomb interaction.  The nonrelativistic Hamiltonian for this system in second quantized notation is
\begin{equation}
\label{H2ndquant}
\begin{split}
& \hat{ H}  =\int d^3r  \,  \hat{q}_i^{\dagger}(\vec{r}) \left (M_{Q} -\frac{\nabla^2}{2 M_Q}   \right )  \hat{q}_i(\vec{r}) \,  +\\
&\frac{\tilde{\alpha}_s}{N_c}  \frac{\lambda_{i j} ^c}{2}  \frac{\lambda^c_{i' j'}}{2} \int d^3r \, d^3r'
\frac{ \hat{q}_i^\dagger(\vec{r})    \hat{q}_{i'}^\dagger(\vec{r'} ) \hat{q}_{j'}(\vec{r'})  \hat{q}_j(\vec{r})  } { |\vec{r}-\vec{r'} | } \; , \\
&{\rm with} \; \tilde{\alpha}_s \equiv \alpha_s N_c \;  ,
\end{split}
\end{equation}
where $\alpha_s=\frac{g^2}{4 \pi}$ is the strong coupling constant (at the appropriate scale for this problem), $\tilde{\alpha}_s$ (the ` t Hooft coupling) is independent of $N_c$,  $M_Q$ is the (common) quark mass, $\hat{q}$ are nonrelativistic fermion field operators with spin and flavor indices suppressed (and implicitly summed over), $\frac{\lambda^c}{2}$ are generators of SU($N_c$) and $i,j,i',j'$ are color indices; there is an implict summation over $c$ and $i, j, i', j'$.

The Hilbert space of physical baryon states consists of color singlet states with $N_c$ quarks.   The color-singlet  nature of the state imposes anti-symmetry with regard to color with each color represented once and only once.  This fact implies that matrix elements of  the Hamiltonian in Eq.~(\ref{H2ndquant}) only receive non-vanishing contributions for terms in which the color of each annihilation operator $\hat{q}$ is identical to the color of one of the creation operators, $\hat{q}^\dagger$, in that term.  Thus in the color Coulomb term there are two types of terms which could contribute: those with $i=j, \, i'=j'$ and those with $i=j', \, i'=j$ :
\begin{equation}
\label{CC}
\begin{split}
&\frac{\tilde{\alpha}_s}{N_c}  \frac{\lambda_{i i} ^c}{2}  \frac{\lambda^c_{i' i'}}{2} \int d^3r \, d^3r'
\frac{ \hat{q}_i^\dagger(\vec{r})    \hat{q}_{i'}^\dagger(\vec{r'} ) \hat{q}_{i'}(\vec{r'})  \hat{q}_i(\vec{r}) } { |\vec{r}-\vec{r'} | } \;  -\\
&\frac{\tilde{\alpha}_s}{N_c}  \frac{\lambda_{i i'} ^c}{2}  \frac{\lambda^c_{i' i}}{2} \int d^3r \, d^3r'
\frac{ \hat{q}_i^\dagger(\vec{r})    \hat{q}_{i'}^\dagger(\vec{r'} ) \hat{q}_{i'}(\vec{r}) \hat{q}_{i}(\vec{r'})  } { |\vec{r}-\vec{r'} | } \;  \, .
\end{split}
\end{equation}
The first term's contribution is at most order $N_c^0$.  Note that it can only  contribute for those generators with nonzero diagonals.  There are only $N_c-1$ of these (each with two nonzero diagonal matrix elements) and thus the sum over $c$ contributes a factor of at most $N_c$ which is canceled by the $1/N_c$ factor out front.  Since the leading order energy is order $N_c$ this term may be dropped.  Moreover, one can use the fact  $\sum_c \lambda_{i i'} ^c \lambda_{i' i} ^c= 2 (1 - \delta_{i i'}/N_c)$ to simplify the color Coulomb term to be
\begin{equation}
\label{CC2}
 - \frac{1}{2} \frac{\tilde{\alpha}_s}{N_c} \int d^3r \, d^3r'
\frac{ \hat{q}_i^\dagger(\vec{r})    \hat{q}_{i'}^\dagger(\vec{r'} ) \hat{q}_{i'}(\vec{r}) \hat{q}_{i}(\vec{r'})  } { |\vec{r}-\vec{r'} | } \;  .
\end{equation}

The anti-symmetry in color implies that the many-body  states  must be symmetric  remaining degrees of freedom---space, spin and flavor.  Thus for n-quark states in the physical state one can suppress the color degrees of freedom and treat the quarks as (spin-1/2)  bosons.   Introduce  bosonic operators, $Q_{fs}(\vec{r})$, which satisfy
\begin{equation}
\begin{split}
[Q_{fs}(\vec{r}),Q_{f's'}(\vec{r'})] &=0 \\
[Q^\dagger_{fs}(\vec{r}),Q^\dagger_{f's'}(\vec{r'})]& =0\\
[Q_{fs}(\vec{r}),Q^\dagger_{f's'}(\vec{r'})] &=\delta_{f,f'}\delta_{s,s'} \delta^3(\vec{r}-\delta{r'}) \; .
\end{split}
\end{equation}
Since the color factors in Eq.~(\ref{CC2}) reduce to simple summing over all of the quarks, one can rewrite the nonrelativistic Hamiltonian as
\begin{equation}
\begin{split}
& \hat{ H}  =\int d^3r  \,  \hat{Q}^{\dagger}(\vec{r}) \left (M_{Q} -\frac{\nabla^2}{2 M_Q}   \right )  \hat{Q}(\vec{r}) \,  - \\
& \,\frac{\tilde{\alpha}_s}{N_c}   \int d^3r \, d^3r'
\frac{  \hat{Q}^\dagger(\vec{r})  \hat{Q}^\dagger(\vec{r'} )    \hat{Q}(\vec{r'})   \hat{Q}(\vec{r})}{ |\vec{r}-\vec{r'} | } \; ,
\end{split}
\label{HB2ndquant}\end{equation}
where we have dropped terms  suppressed in $1/N_c$ and  for simplicity have suppressed explicit reference to the spin and flavor.  In what follows in this section, we will use the bosonic version of the theory.

 Note that at leading order in a heavy quark expansion, as we have here, neither the  spin nor flavor of the quarks  is dynamically relevant--the only role they play is to give the degeneracy of the baryon state.  Accordingly,  in our initial treatment it will be assumed that all quarks are in a single spin and flavor state and the spin and flavor degrees of freedom will be ignored throughout. The issue of degeneracies due to spin and flavor degrees of freedom will be discussed at the end of this section.  In general, even with the neglect of the spin and flavor degrees of freedom,  this is a very complicated quantum mechanical many-body problem.  However, the problem greatly simplifies at large $N_c$.

 The physics underlying this is quite simple.   The interactions between any two quarks makes a small contribution to the energy---of order $1/N_c$.  However, provided that the pairs contribute coherently as in a  mean-field state, the fact that there are of order $N_c^2$ pairs, leads to a contribution of order $N_c$ and, thus, contributes at leading order.  Thus the  key issue is the coherence of the state.   It is apparent from the symmetry under exchange of particles that the optimal coherent state will have each particle in the same quantum state,
\begin{equation}
\begin{split}
& |\psi  \rangle  = \frac {(b^\dagger)^{N_c} }{\sqrt{N_c!}} \, |{\rm vac} \rangle \\
& {\rm with} \; \;  b^\dagger= \int d^3 r \, \psi(\vec{r} ) \, \hat{Q}(\vec{r})
\end{split}
\label{slater}
\end{equation}
where   $\psi(\vec{r})$ is the optimal spatial wave function.
The expectation value of the energy in states of this form is given by
\begin{equation}
		\label{eq:wit}
		\begin{split}
		\langle \psi | \hat{H} | \psi \rangle &= M_{\rm baryon}=  N_c M_Q + N_c \int{d^3r
\frac{|\nabla \psi|^2}{2M_Q}} \\ & - \frac{1}{2} N_c  \tilde{\alpha}_s
\int{d^3 r \, d^3 r' \frac{|\psi(\vec{r})|^2 | \psi (\vec{r'})|^2}{|\vec{r'} - \vec{r'}|}} \; ,\\
&{\rm with} \; \; \int d^3r |\psi(\vec{r})|^2=1 \;.
\end{split}
	\end{equation}
where we have dropped subleading terms in $1/N_c$.
This is the form used in  the mean-field analysis of ref.~\cite{Witten:1979kh}; our expression differs slightly in convention using the standard normalization for $g$ in which $\alpha_s=g^2/(4 \pi)$.  It is convenient to rewrite $N_c^2 \alpha_s$  as $N_c \tilde{\alpha}_s$ which makes clear that all three terms on the right-hand side of Eq.(\ref{eq:wit}) are proportional to $N_c$ .

Minimizing $\langle \psi | \hat{H} | \psi \rangle$ with respect to $\psi(\vec{r})$ gives the leading order result for the mass of the baryon.  We denote this state $ |\psi  \rangle_{\rm  min}$ and the spatial wave function $\psi_{\rm min} (\vec{r})$.  The wave function  $\psi(\vec{r})$  is obtained by minimizing the energy. Starting with Eq.~(\ref{eq:wit}), a straightforward application of the calculus of variations yields an equation for $\psi_{\rm min}(\vec{r})$:	
\begin{equation}
\label{eq:diffeq}
\frac{ 	-\frac{\nabla^2}{ 2 M_Q}\psi_{\rm min}(\vec{r}) -  \tilde{\alpha}_s \psi_{\rm min}(\vec{r})
\int d^3 r' \,
\frac{|\psi_{\rm min}(\vec{r'})|^2}{|\vec{r}-\vec{r'}| }  }{\epsilon}=   \psi_{\rm min}(\vec{r} )\\
	\end{equation}
where $\epsilon$ is a Lagrange multiplier enforcing the normalization.

\subsection{The nature of the mean-field approximation \label{mf}}

States of the form of Eq.~(\ref{slater}) define a variational subspace.  Clearly the nature of variational calculations is such that we can use it to provide a bound on the energy.  However, a question arises as to how well does this state actually approximate the true ground state of the Hamiltonian in Eq.~(\ref{HB2ndquant})   at large $N_c$.   As will be discussed in this subsection, generically one expects   $ |\psi  \rangle_{\rm  min}$ to be a poor approximation to the true ground state, $ |\psi  \rangle_{\rm  exact}$, even in the large $N_c$ limit:   the overlap between  $|\psi  \rangle_{\rm  exact}$ and  $|\psi  \rangle_{\rm  min}$ does not approach unity as $N_c \rightarrow \infty$.   Instead one has
\begin{equation}
1- |{}_{\rm exact}\langle\psi|\psi \rangle_{\rm min}|^2  \sim {\cal O}(N_c^0) \; .
\end{equation}
In practice, it is quite possible that $|{}_{\rm exact}\langle\psi|\psi \rangle_{\rm min}|^2  \ll 1$.  However, despite the fact that  $|\psi  \rangle_{\rm  min}$ is a poor approximation to the true quantum state, ${}_{\rm min} \langle \psi | H | \psi \rangle_{\rm min}$  is a good approximation to the baryon mass  $M_{\rm baryon} ={}_{\rm exact} \langle \psi | H | \psi \rangle_{\rm exact}$: fraction errors are of order $1/N_c$ and thus go to zero at large $N_c$.   This may seem paradoxical at first sight but it has a very simple physical origin: in the exact ground state at large $N_c$,  the probability that {\it all} of the quarks have the wave function $\psi_{\rm min}$ is small while the probability is very high that {\it most} (that is, all except a fraction  of order $1/N_c$ ) of the quarks have this wave function and thus dominate the energy functional.

To understand why this is so, it useful to decompose the Hamiltonian into a one-body mean-field Hamiltonian and  correction terms:
\begin{equation}
\label{decomp}
\begin{split}
\hat{H}&=\hat{H}_{\rm mf} + \Delta \hat{H}\\
\hat{H}_{\rm mf} & \equiv \hat{T} + \hat{V}_{\rm mf} -C\\
\Delta \hat{H} & \equiv \hat{V} -\hat{V}_{\rm mf}+ C\\
\end{split}
\end{equation}
with
\begin{equation}
\begin{split}
\label{decomp1}
\hat{T}& \equiv -  \frac{Q^\dagger(\vec{r}) \nabla^2Q^\dagger(\vec{r}) }{2 M_Q} \\
\hat{V}& \equiv -\left ( 1+\frac{1}{N_c} \right ) \frac{\tilde{\alpha}_s}{N_c}   \int d^3r \, d^3r'  \frac{  \hat{Q}^\dagger(\vec{r}) \hat{Q}(\vec{r})  \;  \hat{Q}^\dagger(\vec{r'} )  \hat{Q}(\vec{r'})   } { |\vec{r}-\vec{r'} | }\\
\hat{V}_{\rm mf}& \equiv  -  \tilde{\alpha}_s Q^\dagger(\vec{r}) \left (\int d^3 r' \,  \frac{|\psi_{\rm min}(\vec{r'})|^2}{|\vec{r}-\vec{r'}| } \right ) Q^\dagger(\vec{r}) \\
C& =-  \frac{   \left (N_c+\frac{1}{N_c}   \right ) \tilde{\alpha}_s }{2}  \int d^3 r d^3 r' \,  \frac{|\psi_{\rm min}(\vec{r'})|^2|\psi_{\rm min}(\vec{r'})|^2}{|\vec{r}-\vec{r'}| }
\end{split}
\end{equation}
The form of the one-body mean-field Hamiltonian is seen in the one-body Schr\"odinger equation of Eq.~(\ref {eq:diffeq}).  It is easy to see  that $ {}_{\rm min}\langle \psi | \hat{V} |\psi \rangle_{\rm min} =    {}_{\rm min}\langle \psi | \hat{V}_{\rm mf}  |\psi \rangle_{\rm min} \left (\frac{1}{2} - \frac{1}{2N_c^2} \right)$ and $C$ is chosen so that  $ {}_{\rm min}\langle \psi | \Delta \hat{H} |\psi \rangle_{\rm min} = 0$.  The eigenstates of the one-body mean-field Hamiltonian,  $\hat{H}_{\rm mf}$, provide a complete basis of states.  By construction, $|\psi \rangle_{\rm min}$ is the ground state of  $\hat{H}_{\rm mf}$.  Excited states of   $\hat{H}_{\rm mf}$ are  $n$-particle-$n$-hole excitations relative to  $|\psi \rangle_{\rm min}$.   In the large $N_c$ limit, the excitation energy of a generic $n$-particle-$n$-hole  state is independent of $N_c$.

While  $|\psi \rangle_{\rm min}$ is the ground state of  $\hat{H}_{\rm mf}$ it is clearly not the ground state of  $\hat{H}$.  The presence of $\Delta \hat{H}$  in $\hat{H}$ introduces correlations in the form of $n$-particle-$n$-hole state contributions to the ground state.  At first blush it may seem reasonable to regard $\Delta \hat{H}$ as a perturbation and use perturbation theory to compute these correlations.  In implementing such a calculation it is useful to understand the $N_c$ scaling of matrix elements of $\Delta \hat{H}$ between various $n$-particle-$n$-hole states (including the 0-particle-0-hole state $|\psi \rangle_{\rm min}$):
\begin{equation}
\label{ME}
\begin{split}
\langle {\rm( n\pm1) p-(n\pm 1) h} | \Delta \hat{H}| {\rm n p-n h} \rangle  &\sim N_c^{-\frac{1}{2}} \\
\langle {\rm( n\pm2) p-(n\pm 2) h} | \Delta \hat{H}| {\rm n p-n h} \rangle  & \sim N_c^{0}\\
\langle {\rm( n \pm m) p-(n \pm m) h} | \Delta \hat{H}| {\rm n p-n h} \rangle & =0 \; \; {\rm for} \; m>2
\end{split}
\end{equation}
At first order in perturbation theory, there is no correction due to $\Delta \hat{H}$.  At second order,
\begin{equation}
\label{enper}
\begin{split}
\Delta M_{\rm baryon} & = \sum_{n \in 1p-1h} \frac{|\langle n | \Delta \hat{H}| \psi\rangle_{\rm min}|^2}{E_n^{\rm mf}-E_{0}^{\rm mf} } \\ &+ \sum_{n \in 2p-2h} \frac{|\langle n  | \Delta \hat{H}| \psi\rangle_{\rm min}|^2}{E_n^{\rm mf}-E_{0}^{\rm mf} }
\end{split}
\end{equation}
where the superscript mf indicates that these are the energies in the unperturbed mean-field Hamiltonian.  From Eq.~(\ref{ME}) it is apparent that the contribution from the one-particle-one-hole states are of order $1/N_c$ while the contributions from the two-particle-two-hole states are of order $N_c^0$.  Since the mean-field calculation gives a mass of order $N_c^1$, this correction is down by a factor of order $N_c^{-1}$.

The correction to the energy in perturbation theory is parametrically small compared to the leading order contribution.  This is what one would expect if the system was in the regime of validity of perturbation theory.  However, if one looks at the correction to the state, one sees that the dominant  (2-p-2-h) contribution is given by
\begin{equation}
\label{stateper}
\Delta |\psi\rangle  =\sum_{n \in 2p-2h} |n \rangle  \frac{\langle n  | \Delta \hat{H}| \rangle_{\rm min}}{E_n^{\rm mf}-E_{0}^{\rm mf} } \sim N_c^0 \; .
\end{equation}
Clearly, low-order perturbation theory cannot  be justified on parametric grounds at large $N_c$; corrections to the state are generically not small.

The failure of low-order perturbation theory in $\Delta \hat{H}$  seen in Eq.~(\ref{stateper}) implies that the perturbative computation of the mass shift in  Eq.~(\ref{enper}) is  unreliable.   This presents a major challenge in any attempt to compute subleading correction to the mass in the $1/N_c$ expansion; these include contributions due to perturbations due to $\Delta \hat{H}$.   However, for the purposes of extracting the leading order behavior,  it is not necessary to  compute the correction accurately: the relevant issue is the $N_c$ scaling of the contribution to the mass due to the existence of $\Delta \hat{H}$.

Fortunately,  perturbation theory does  capture the $1/N_c$ scaling properly.   To see why, one can imagine doing the following artificial calculation. Consider the eigenspectrum of
\begin{equation}
\hat{H} _\lambda = \hat{H}_{\rm mf} + \lambda \Delta \hat{H}
\end{equation}
 where $\lambda$ is a parameter.   The actual problem of interest corresponds to $\lambda=1$.  In this artificial problem  $\Delta M_{\rm baryon}$ is a function of both $\lambda$ and $1/N_c$.     Let us consider what happens for small $\lambda$.  In this case we can justify perturbation theory in $ \lambda \Delta \hat{H}$:
 \begin{equation}
 \Delta  M_{\rm baryon} =\sum_{j=0}^\infty  \lambda^j m_j \; .
 \end{equation}
 Using the scaling rules in Eq.~(\ref{ME}), it is easy to show that each of the coefficients $m_j$ scales as $N_c^0$ at leading order in the $1/N_c$ expansion.  Expanding the coefficients  $m_j$  in a series of $1/N_c$ yields a double expansion:
 \begin{equation}
 \Delta M_{\rm baryon} =\sum_{j=0}^\infty  \sum_{k=0}^\infty  \lambda^j  N_c^{-k}   \,m_{j k} \;
 \end{equation}
 where the coefficients $m_{j k}$ are independent of $\lambda$ and $N_c$.
 Providing that the double expansion is well behaved, one can switch the order of summation yielding
 \begin{equation}
 \begin{split}
 \Delta M_{\rm baryon} & =\sum_{j=0}^\infty   N_c^{-j}   M_j(\lambda)   \;  \; {\rm with} \\
  M_j(\lambda) & \equiv \sum_{k=0}^\infty  \lambda^k \,m_{jk} \; .
  \end{split}
 \end{equation}
 In effect, in introducing $M_j$ we have resummed the $\lambda$ expansion to get the contribution to all orders.  The leading order term in the $1/N_c$ expansion thus scales as $N_c^0$ for any value of $\lambda$.      Extrapolating to $\lambda=1$ shows that corrections to $M_{\rm baryon}$ due to the presence of $ \Delta \hat{H}$ are generically of absolute order $N_c^0$ which is of relative order $1/N_c$ compared to the leading order contribution.   The argument given here does have a loop-hole in that it depends on the technical assumption that the double expansion is well behaved. We have no reason to doubt that this is the case.

Thus, the mean-field approach advocated by Witten\cite{Witten:1979kh} appears to be valid as a technique for obtaining the leading order value of the mass in a $1/N_c$ expansion.  However, it does not give the leading order wave function.  Moreover,  the  $1/N_c$ expansion is  not equivalent to a perturbative expansion in $\Delta \hat{H}$.

The apparent discrepancy  between mean-field theory providing  an accurate computation of the energy while  yielding an inaccurate wave function is easy to understand.  At a physical level it is apparent in the exact ground state at large $N_c$,  the probability that {\it all} of the quarks have the wave function $\psi_{\rm min}$ is small; this explains the breakdown of mean-field theory for the wave function.  At the same time
only a small fraction of  the quarks (of order $1/N_c$ )  are not in the mean-field single-particle level.  Thus the mean-field quarks dominate the energy functional.  From this physical picture, it is immediately apparent that one can accurately compute observables beyond the energy in mean-field theory.  In particular, any $n$-body operator with $n \ll N_c$ can be calculated; the expected error will be of relative order $n/N_c$.

One final comment about the nature of this mean-field approximation.  The analysis above was in the framework of the effective bosonic theory.   Of course, the result is the same if one works in the original fermionic  theory.  In that case, instead of using the coherent state of Eq.~(\ref{slater}) one would use a Slater determinant.  From the form of the interaction term in Eq.~(\ref{CC2}), it is apparent that the term which contributes is the exchange interaction and not the direct interaction.  This means that the contribution comes from the Fock  term ({\it i.e.}, the exchange interaction) and not the Hartree term ({\it i.e.}, the direct interaction).  Thus, the description in
ref.~\cite{Witten:1979kh} of the approximation as a Hartree approximation  is something of  a misnomer.  It is important to keep this in mind when generalizing the treatment to the problem of baryonic matter, as is done later in this paper.

\subsection{Parametric dependence}

Before  explicitly implementing a variational calculation to find the baryon mass, it is useful to understand the parametric dependence of the mass on $N_c$, $\tilde{\alpha}_s$ and $M_Q$.  The $N_c$ dependence is trivial: the mass is clearly proportional to $N_c$.   More generally, the mass of the baryon can be seen to be given by $
M_{\rm baryon}= N_c M_Q \left (1 -{\rm const} \, \tilde{\alpha}_s^2 \right )$.
This can be shown via a simple scaling argument.  Note that the right-hand side of  Eq.(\ref{eq:wit}) has three terms: a quark mass term, a quark kinetic term, and an attractive interaction term due to color Coulomb interactions.  One can ask how these three terms scale with the spatial size of the wave function?  Introduce a length scale parameter $R$ with
\begin{equation}
\psi(\vec{r})= R^{-3/2} f_0(\vec{r}/R)
\label{psiform}\end{equation}
where   $f_0(\vec{y})$  is a reference normalized functional form which characterizes the {\t shape} of the wave function; $y$ is dimensionless and the {\it size} of the wave function is thus characterized by $R$.
\begin{equation}
		\label{eq:varalform}
		\begin{split}
		\langle \psi | H | \psi \rangle  &=  N_c (M_Q + t + v) \\ & = N_c \left  (M_Q + \frac{ {\cal T}(f_0)}{R^2 M_Q}  -   \tilde{\alpha}_s \frac{ { \cal V}(f_0)}{R}  \right )\\
		 & \;  {\cal T}(f_0)\equiv \int d^3 y \frac{|\vec{\nabla}_y f_0(\vec{y})|^2}{2} \\
		& \; {\cal V}(f_0)\equiv \int d^3 y  \, d^3 y' \frac{| f_0(\vec{y})|^2 | f_0(\vec{y'})|^2}{2 |\vec{y}-\vec{y'}|} \;
		\end{split}
\end{equation}		
 $t $ and $v$ are the kinetic and interaction terms, and   ${\cal T}$  and ${\cal V}$ are dimensionless functionals of the reference function $f_0$.    Suppose $\psi_0$ is the wave function that minimizes the  energy.  In that case we know that $\partial_R \langle \psi | H | \psi \rangle =0$.  From this we deduce that
 \begin{equation}
 \label{scaleresult}
 \begin{split}
 R& =\frac{1}{\alpha_s M}  \frac{2 {\cal T}(f_0)}{{\cal V}(f_0)} \\
 t &= -\frac{v}{2}=   \tilde{\alpha}_s^2 M_Q \frac{ {\cal V}^2  (f_0)}{4 {\cal T}(f_0)} \\
 \langle \psi | H | \psi \rangle  & = M_{\rm baryon} = N_c  M_Q \left (1 - \tilde{\alpha}_s^2  \frac{ {\cal V}^2  (f_0)}{4 {\cal T}(f_0)}  \right ) \; .
\end{split}
\end{equation}
Thus $N_c M_Q \tilde{\alpha}_s^2  \frac{ {\cal V}^2  (f_0)}{4 {\cal T}(f_0)}$ can be regarded as the binding energy of the heavy quarks.  Note that an overall dilation of $f_0$ is innocuous: it changes the equilibrium value of $R$, ${\cal T}$ and  ${\cal V}$ while keeping $ \langle \psi | H | \psi \rangle$ unchanged.   A direct comparison of Eq.~(\ref{eq:diffeq})    with Eqs.~(\ref{eq:varalform})  and  (\ref{scaleresult}) shows that
\begin{equation}
\epsilon=t+2v = -\frac{3 \tilde{\alpha}_s^2 M_Q }{4} \frac{{\cal V}^2}{\cal T} \; .
\end{equation}

Thus the problem of finding the mass of the baryons in this limit reduced to determining a single function $f_0$ from which ${\cal T}$ and ${\cal V}$ and, hence, the mass of the baryon can be computed.  The functional form  for $f_0$  does not depend on  $\tilde{\alpha}_s$ or $M_Q$ and thus neither do ${\cal T}$ and ${\cal V}$.

\subsection{A variational calculation}

Equation~({\ref{eq:wit}) is a nonlinear integro-differential equation.  One standard strategy in obtaining  an approximate numerical solution   to such equations is iterative.  An alternative approach which is pursued here is to consider a variational space of trial wave functions.  A wave function in such a space  is specified by some finite set of  parameters $b_1,b_2 \ldots, b_n$.    Evaluating the energy in Eq.~(\ref{eq:wit}),  using states in the variational space and minimizing with respect to the parameters   $b_1,b_2 \ldots, b_n$ will yield an optimal wave function for the space of trial wave functions.  If the space is rich enough to include the exact solution of  Eq.~(\ref{eq:diffeq}), then the wave function  obtained by minimizing with respect to $b_1,b_2 \ldots,b_n $ will necessarily be a solution of Eq.~(\ref{eq:diffeq}).  If  the trial space does not contain the exact solution but does contain states with a very large overlap with the exact solution, one expects that  solution obtained via minimization in the trial space to approximate the exact solution with high accuracy.  The virtue of this approach is that one finds explicit analytic expressions for the wave functions---albeit only approximate ones.   Moreover, there is considerable experience indicating that even quite modest variational spaces can yield remarkably accurate results for energies.

One can represent a function in  the trial class of normalized  dimensionless functions  as $f_0^{(b_1, b_2, \ldots b_n)}(\vec{y})$  so that  the wave function is given by $\psi^{(b_1, b_2, \ldots b_n)}(\vec{r})= R^{-3/2} f_0^{(b_1, b_2, \ldots b_n)}(\vec{r}/R)$.   The expectation value of the energy in this state is given by
\begin{equation}
\label{evar}
\begin{split}
&E(R, b_1, b_2, \ldots b_n)= \langle \psi^{(b_1, b_2, \ldots b_n)}| H | \psi^{(b_1, b_2, \ldots b_n)} \rangle =\\  &N_c \left  (M_Q + \frac{ {\cal T}(f_0^{(b_1, b_2, \ldots b_n)})}{R^2 M_Q}  -   \tilde{\alpha}_s \frac{ { \cal V}(f_0^{(b_1, b_2, \ldots b_n)})}{R}  \right ) \; .
\end{split}
\end{equation}
The optimal values for the coefficients are obtained from minimizing $E(R, b_1, b_2, \ldots b_n)$ with respect  to $R$ and the $b_i$.    Extremizing $E(R, b_1, b_2, \ldots b_n)$  yields the following set of coupled equations:
\begin{equation}
\label{cond}
\begin{split}
\frac{\partial \log {\cal T}(f_0^{(b_1, b_2, \ldots b_n)})}{\partial b_1}& = 2 \frac{\partial \log {\cal V}(f_0^{(b_1, b_2, \ldots b_n)})}{\partial b_1} \\
\frac{\partial \log {\cal T}(f_0^{(b_1, b_2, \ldots b_n)})}{\partial b_2}& = 2 \frac{\partial \log {\cal V}(f_0^{(b_1, b_2, \ldots b_n)})}{\partial b_2} \\ & \vdots \\ \frac{\partial \log {\cal T}(f_0^{(b_1, b_2, \ldots b_n)})}{\partial b_n}& = 2 \frac{\partial \log {\cal V}(f_0^{(b_1, b_2, \ldots b_n)})}{\partial b_n}  \; .\\
\end{split}
\end{equation}
For any particular class of functions, Eq.~(\ref{cond}) is simply a set of algebraic equations and is amenable to standard numerical solutions.  It is easy to verify that a particular solution is a (local) minimum as opposed to a maximum or saddle point.

To implement this scheme in practice one needs to choose a space of   functions which is rich enough to closely approximate the exact solution.  There is a strong theoretical  prejudice that the exact solution should be spherically symmetric.   While it is certainly possible for mean-field solutions to break symmetries, in the present circumstance it seems clear that a spherical shape both minimizes the kinetic energy and maximizes the magnitude of attractive color Coulomb potential energy.    Thus, our trial class of functions consists of spherically symmetric functions.

One useful way to construct a trial class of functions is to begin with  a complete set of spherically symmetric functions, $\{ \phi_0(y), \phi_1(y), \phi_2(y), \ldots \}$, which satisfy an orthonormality condition, $\int_0^\infty d y \, \phi^*_j(-) \phi^*_k(y) = \delta_ {j,k}$.   The exact form of $f_0$ can always be written as a normalized superposition of the $\phi_j$.  A truncation of the sum in the superposition yields a useful trial class of functions:
\begin{equation}
f_0^{(b_1, b_2, \ldots b_n)}(\vec{y})= ( 1- \sum_{j=1}^n | b_j|^2  )^{\frac{1}{2}}  \phi_0(y)+ \sum_{j=1}^n b_j \phi_j(y)  \; .
\end{equation}
This form is viable provided that the shape of  an exact solution to Eq.~({\ref{eq:diffeq}) has a nonvanishing overlap with $\phi_0$.  By truncating the expansion at a different value of $n$ one can test numerical   convergence.     A particularly useful complete set of functions for this construction is 
\begin{equation}
\label{ho}
\phi_j(\vec{y})  =
 \frac{ (-1)^j}{y \sqrt{(2j+1)!}}  \left(\frac{2}{\pi} \right)^{\frac{3}{4}}   \left ( y + \frac{1}{2} \frac{d}{d y} \right)^{2 j} \left (y e^{-y^2} \right ) \; ;
 \end{equation}
these are the s-wave eigenfunctions of the harmonic oscillator (up to an irrelevant scale factor) and a sign convention with $\phi_j(0)$ positive.   These functions have the virtue that  an  analytic expression for energy in terms of the $b_j$  can be computed straightforwardly. Moreover, with this  trial class of functions the energy converges quite rapidly as additional states are included into the sum.

\begin{table}
\label{TV}
\caption{This table shows the numerical convergence of the coefficients ${\cal T}$ and ${\cal V}$  defined in Eq.~(\ref{eq:varalform}) for the minimum energy configuration computed in a truncated basis.  The functions are given in Eq.~(\ref{ho}).  The combination $\frac{{\cal V}^2}{4 \cal T}$ determines the binding energy of the heavy quarks.}

\begin{center}

\begin{tabular}{| c | c | c | c | c | }

\hline & & &\\
basis 	& $\frac{{\cal V}^2}{4 \cal T}$ &  ${\cal T}$ &  ${\cal V}$   \\
functions &  & & \\
\hline
$\phi_0 $	                                      & 0.053052 &	1.5000 &	0.56419   \\			
\hline$\phi_0, \phi_1,\phi_2 $	           & 0.054198    & 1.4293  &0.55666   \\
\hline $\phi_0, \phi_1,\ldots , \phi_4  $   &0.054252	&1.4067  &0.55250 	  \\
\hline$\phi_0, \phi_1,\ldots , \phi_6  $	   &0.054256   &1.3941   &0.55006	\\
\hline
	
\end{tabular}
\end{center}
\end{table}

We have minimized the energy function using truncations with up to seven basis states, ($\phi_0, \ldots \phi_6$).   The results for ${\cal T}$ and ${\cal V}$  are given in Table I.  It is apparent that the quantity $\frac{{\cal V}^2}{4 \cal T}$ which determines the binding energy of the heavy quarks is well converged and is  given by 0.05426 up to four significant  figures.  Thus in the heavy quark and large $N_c$ limits the mass of the baryon is well approximated by
\begin{equation} \label{baryon}
 M_{\rm baryon}  \approx N_c  M_Q \left (1 - 0.05426 \,  \tilde{\alpha}_s^2  \right ) \, .
 \end{equation}
The  wave function at this level of truncation is determined by the coefficients $b_1 \ldots b_6$  which, to good numerical approximation, are given by
\begin{equation}
\begin{split}
& b_1\approx -0.0476828  \, , \;  b_2\approx 0.0815798 \, , \; b_3\approx -0.00786028 \, , \\  & b_4\approx 0.0136048 \, ,\; b_5\approx -0.00159968 \, , \; b_6\approx 0.00283365  \;
   \end{split}
   \end{equation}
 with the scale factor $R$ given by
 \begin{equation}
R=\frac{1}{\alpha_s M}  \frac{2 {\cal T}(f_0)}{{\cal V}(f_0)}  \approx  \frac{5.06907}{\alpha_s M}  \, .
\end{equation}
Combing these yields the following approximate wave function:
\begin{equation}
\label{approxwf}
\begin{split}
\psi_{\rm min}(\vec{r})& \approx {0.0876207}{ (\tilde{\alpha}_s M_Q)^{3/2} } \exp \left (-0.0389173  \bar{r} ^2  \right)\\
\times & \left  (0.755925 - 0.00856005 \bar{r} ^2 +
   0.000289408  \bar{r} ^4  \right . \\
  & - 3.69934 \, 10^-{6} \bar{r} ^6 + 3.58608 \, 10^{-8}   \bar{r} ^8 \\
   &\left .- 1.65987 \, 10^{-10} \bar{r} ^{10} +  3.64193 \,10^{-13}  \bar{r} ^{12} \right)\\
   &{\rm with} \; \;  \bar{r} \equiv r \tilde{\alpha}_s M_Q
\end{split}
\end{equation}
The  wave function is plotted in Fig.~\ref{fig:fig1}.

\begin{figure}
\includegraphics[width=2.9in]{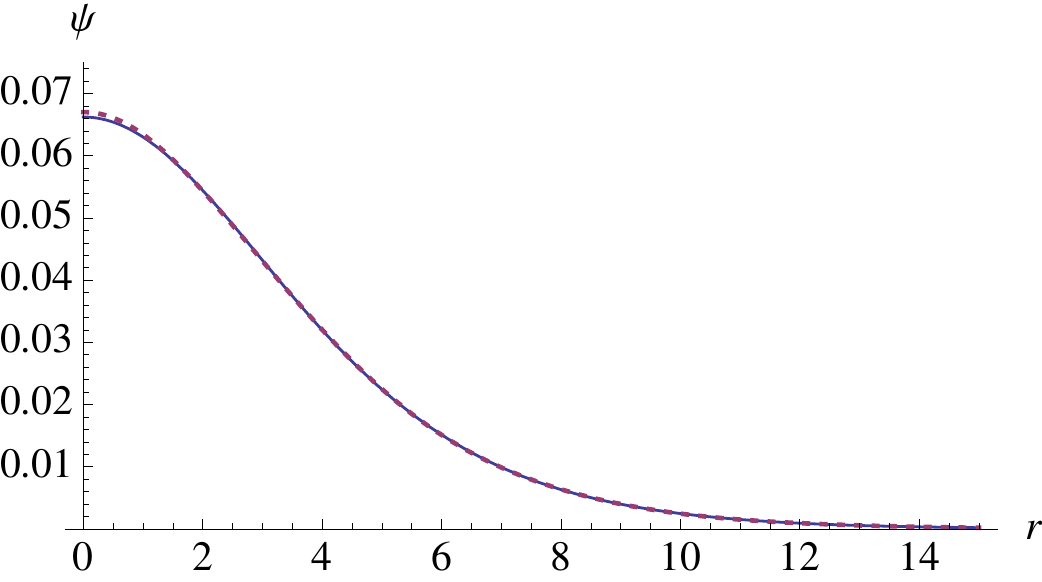}
\caption{The variational wave function of Eq.~(\ref{approxwf}) is  plotted against $r$ in the solid curve.   $\psi$ is given in units of $(\tilde{\alpha}_s M_q )^{3/2}$ and $r$ is given in units of $(\tilde{\alpha}_s M_q )^{-1}$  .  The short-dashed curve represents the left-hand side of  Eq. (\ref{eq:diffeq}) using the variational wave function of Eq.~(\ref{approxwf}).  For the exact wave function, the two coincide.}
\label{fig:fig1}
\end{figure}

It is clear from Table I that the variational procedure  converges rapidly yielding an accurate expression for the mass.  Of course, it is the nature of variational approximations that energies are computed more accurately than the wave functions: first order difference between the approximate and exact wave functions do not contribute to the energy.   A useful way to test the accuracy of the variational wave function itself is to compare the left-hand side of  Eq.~(\ref{approxwf}) with $\psi$; for the exact wave function the difference is zero.   This comparison is made in Fig.~\ref{fig:fig1} where the short-dashed curve represents the left-hand side of  Eq.~(\ref{approxwf}) using the variational wave function and the solid curve is  the variational wave function.  Note that overall the two curves  are quite close everywhere.  The largest difference between the two is at the origin and amounts to a 1\% discrepancy.  Elsewhere the discrepancy is quite small for the entire region where the wave function is substantial. For typical observables this is the region of relevance and thus we conclude that we can accurately compute typical matrix elements with high precision---at least for a world with $N_c$ and $M_Q$  very large.

Note that the form of all of the wave functions in the variational space used here  drop off at very large distances quite rapidly---like a power law times a Gaussian.  As will be shown below, the exact wave function drops off as a power law times a decaying exponential,  which is much slower than the states in the variational space.  This means that at very long distances the variational wave function is much smaller than the actual solution of  Eq.~({\ref{eq:wit}).   Fortunately the variational space used is rich enough so that, although the wave function is inaccurate at asymptotically large distances, it is quite accurate over the domain containing the vast bulk of the baryon's probability.   Hence the variational wave function is capable of describing virtually all of the standard baryon observables of interest such as the mass, the baryon charge radius or the form factor in the range where $Q^2 \sim (M_Q \alpha)^2$.  However, if one focuses on observables  which are dominated by the behavior at a very large distance, the wave function in Eq.~(\ref{approxwf}) is inadequate.

Fortunately, it is possible to determine the long distance behavior of the wave function accurately.  One can start with the variational wave function Eq.~(\ref{approxwf}) and determine the form of the correct asymptotic of the solution Eq.~({\ref{eq:wit}).  One can match this onto  the variational wave function in a domain where both the asymptotic form and the variational wave function are accurate.    The asymptotic form is found easily: for large $r$,  $\int d^3 r' \,  \frac{ |\psi_{\rm min}(\vec{r'})|^2}{|\vec{r}-\vec{r'}| }  \rightarrow \frac{1}{r}$ and Eq.~({\ref{eq:wit}) asymptotically is identical to the Schr\"odinger equation for a Coulomb potential.  Of course the solutions are not hydrogenic---even asymptotically---since the eigenenergy, $\epsilon$, does not correspond to those of hydrogen atom.  Using this asymptotic form of the potential and solving the Schr\"odinger equation yields an asymptotic wave function of
   \begin{equation}
    \label{eq:longrange}
    \begin{split}
    &\psi_{\rm Asy} (r)=(M_Q \tilde{\alpha}_s)^{3/2} \bar{\psi}_{\rm Asy}(\bar{r})  \; \; {\rm with}\\
    &\bar{\psi}_{\rm Asy}(\bar{r}) = \frac{\beta}{\bar{r}}e^{-\bar{r}\sqrt{-2\bar{\epsilon}}}\;{\rm U}(\frac{\bar{\epsilon}}{\sqrt{-2\bar{\epsilon}^3}},\;0,\;2\bar{r}\sqrt{-2\bar{\epsilon}}) \, ,\\
    &\bar{r} = (M_Q \tilde{\alpha}_s) \;  , \; \;  \; \; \bar{\epsilon}=\frac{\epsilon}{M_Q \tilde{\alpha}_s^2} \, ;\;
    \end{split}
    \end{equation}
U is the confluent hypergeometric function of the second kind. The constant,  $\beta$, can be fixed by matching this asymptotic form to the variational solution.  The matching needs to be done in a regime where $r$ is large enough so that the corrections to the asymptotic form of Eq.~({\ref{eq:wit}) are negligibly small while simultaneously being sufficiently small so that errors in the variational wavefuntion due to its wrong asymptotic behavior   are also negligible.    By comparing the $r$ dependence of both the variational and asymptotic wave functions we found that these conditions were well satisfied in the region $7<\bar{r}<11$.  We chose to match at $\bar{r}=9$ and found
\begin{equation}
\beta \approx .118
\label{beta}
\end{equation}
We estimate the error in the determination of $\beta$ to be less than .5\%.

\subsection{The baryon form factor}

One of the most important quantities characterizing a baryon is its form factor.  For a nonrelativistic system such as the one being considered here it is defined by
\begin{equation}
\label{FF}
\begin{split}
G(q^2) & = \langle \vec{p'}| \hat{\rho}(\vec{0})|\vec{p} \rangle \\
q^2 & \equiv \left (\vec{p'}-\vec{p} \right )^2  \\
 \hat{\rho}_b (\vec{x})&\equiv \frac{\hat{q}^\dagger(\vec{x}) \hat{q}(\vec{x}) }{N_c}
\end{split}
\end{equation}
where $|\vec{p}\rangle$ is a baryon state of good momentum defined with the normalization $\langle \vec{p'}|\vec{p}\rangle=(2 \pi)^3 \delta^3(\vec{p} -\vec{p'})$ and $\hat{q}$ is the quark field operator.  $G$ is essentially the relativistic  Sachs form factor $G_E$ in the regime where the momentum transfer is much less than the baryon mass.    For the equivalent  bosonic theory that we are using here, the density operator may be written as $ \hat{\rho}_b (\vec{x})\equiv \frac{\hat{Q}^\dagger(\vec{x}) \hat{Q}(\vec{x}) }{N_c} $.

 One typically expects that in mean-field  models that break translational invariance, the form factor is given by the Fourier transform of density in the center of mass:
\begin{equation}
\label{FFFT}
G_{\rm mf}(q^2) =\int d^3 x \, {\rm e}^{i \vec{q} \cdot \vec{x}}  |  \psi_{\rm min}(\vec{x})|^2 \; .
\end{equation}
It is useful to verify that  Eq.~(\ref{FFFT}) does indeed follow from the  mean-field approximation,  and that it is justified at large $N_c$ for heavy quarks. In doing so one can  determine the parametric dependence of the corrections to this result.

As noted in Subsec. \ref{mf}, one expects the mean-field formalism to be valid at large $N_c$ for few-body operators.  As $\hat{\rho}$ is a one-body operator it appears that mean-field theory should be appropriate.  There is a subtlety, however,  in that  the mean-field state breaks translational invariance and thus is not an eigenstate of momentum; accordingly it cannot be used directly in Eq.~(\ref{FFFT}).  In order to compute the form factor it is necessary to decompose the mean-field state as an integral over momentum eigenstates
\begin{equation}
\begin{split}
|\psi \rangle_{\rm min}&= \int \frac{d^3 p}{(2 \pi)^3}  A(\vec{p}) | \vec{p}  \rangle_{\rm mf}\\
A(\vec{p}) | \vec{p}  \rangle_{\rm mf}&\equiv  \int d^3 x \, {\rm e}^{i(\vec{p} - \hat{\vec{P}} ) \cdot \vec{x}} |\psi \rangle_{\rm min} \\
\hat{\vec{P}}& \equiv \int d^3 x \, Q^\dagger(\vec{x}) \left ( - i \vec{\nabla} \right  ) Q(\vec{x}) \; ;
\end{split}
\label{A}
\end{equation}
$ |\vec{p}  \rangle_{\rm mf} $ is the (normalized) state of good momenta obtained by projecting from the mean-field state and $\hat{\vec{P}}$ is the momentum operator.  It is straightforward to determine the parametric dependence of $A(\vec{p})$ on $N_c$, $M_Q$ and $\tilde{\alpha}_s$:
\begin{equation}
A(\vec{p}) =\frac {    \tilde{A} \left ( \frac{\vec{p}}{  \sqrt{N_c} \tilde{\alpha}_s M_Q     }  \right )  }{  \left (\sqrt{N_c} \tilde{\alpha}_s M_Q \right )^\frac{3}{2} }
\label{AScale}\end{equation}
where the functional form of $  \tilde{A}$ is independent of $N_c$, $M_Q$ and $\tilde{\alpha}_s$.

Consider $ \int d^3 x \, {\rm e}^{i \vec{q} \cdot \vec{x}}   {}_{\rm min} \langle \psi   |
\hat{\rho}_b(\vec{x}) \psi \rangle_{\rm min} $.   On the one hand, using Eq.~(\ref{slater}) and the definition of $\hat{\rho}_b$ this is easily seen to be given by
\begin{equation}
 \int d^3 x \, {\rm e}^{i \vec{q} \cdot \vec{x}}   {}_{\rm min} \langle \psi   |
\hat{\rho}_b(\vec{x}) \psi \rangle_{\rm min}  = \int d^3 x \, {\rm e}^{i \vec{q} \cdot \vec{x}}  |  \psi_{\rm min}(\vec{x})|^2 \; .
\end{equation}
On the other hand, using Eq.~(\ref{A}) and the fact that $\rho(\vec{x})= e^{i  \hat{\vec{P}}\cdot \vec{x}} \rho(\vec{0}) e^{-i  \hat{\vec{P}}\cdot \vec{x}}$, it can be written as
\begin{equation}
\begin{split}
 &\int d^3 x \, {\rm e}^{i \vec{q} \cdot \vec{x}}   {}_{\rm min} \langle \psi   |
\hat{\rho}_b(\vec{x}) | \psi \rangle_{\rm min}  =  \\
& \int d^3 x \, \frac{d^3 p'}{(2 \pi)^3} \, \frac{d^3 p }{(2 \pi)^3}\,  {\rm e}^{i (\vec{q}+\vec{p}'-\vec{p}) \cdot \vec{x}}  A^*(\vec{p}') A(\vec{p}) \,   {}_{\rm mf}\langle p  | \hat{\rho}_b(\vec{0})   |  \vec{p}  \rangle_{\rm mf}\\
&=  G_{\rm mf} (q^2) \int \frac{d^3 p }{(2 \pi)^3}\,  A^*(\vec{p}) A(\vec{p}+\vec{q})\\
&= G_{\rm mf} (q^2)\int \frac{d^3 \tilde{p} }{(2 \pi)^3}\,  \tilde{A}^*(\vec{\tilde{p}}) \tilde{A} \left (\vec{\tilde{p}}+\vec{q}\left (\sqrt{N_c} \tilde{\alpha}_s M_Q \right )^{-1} \right) \; .
\end{split}
\label{FF2}\end{equation}
where $G_{\rm mf}$ is the baryon form factor defined in Eq.~(\ref{FF}) evaluated using the mean-field state $|\psi\rangle_{\rm min}$ and $\tilde{A}$ is the dimensionless form introduced in Eq.~(\ref{AScale}).   Note  that at large $N_c$, with $q$ independent of $N_c$, $\vec{q} \left ( \sqrt{N_c} \tilde{\alpha}_s M_Q \right )^{-1} $ goes to 0;   dropping it in the last form of Eq.~(\ref{FF2}) yields  $ G_{\rm mf} (q^2)$  since the  integral of $| \tilde{A}|^2$  is unity.   Thus, at large $N_c$ we have $ G_{\rm mf} (q^2) =  \int d^3 x \, {\rm e}^{i \vec{q} \cdot \vec{x}}   {}_{\rm min} \langle \psi   |
\hat{\rho}_b(\vec{x}) | \psi \rangle_{\rm min}$ which immediately yields Eq.~(\ref{FFFT}).

At finite $N_c$ there are correction terms to Eq.~(\ref{FFFT}) since $\left ( \sqrt{N_c} \tilde{\alpha}_s M_Q \right )^{-1}  \neq 0$.    It is easy to see that parametrically they are formally  of relative order $q^2   \left ({N_c} \tilde{\alpha}_s^2 M_Q^2 \right )^{-1} $.    The interesting kinematic region is for $Q^2 \sim  \tilde{\alpha}_s^2 M_Q^2$ since this is the characteristic scale over which the form factor varies.  Thus, in this region there is a correction of relative order $1/N_c$  to  Eq.~(\ref{FFFT})  from the approximation given above.  There are  other   corrections of the same order due to the fact that the mean-field wave  function is not exact.

We have evaluated the form factor using  Eq.~(\ref{FFFT}) and the wave function in Eq.~(\ref{approxwf}).  The result is plotted in Fig.~\ref{fig:fig2}.  It is of interest to consider derivatives of  the form factor with respect to $q^2$ evaluated at $q^2=0$ since these are related to moments of the baryon density distribution:
\begin{equation}
\begin{split}
\langle r^2 \rangle_{\rm baryon}  &= -6 \left . \frac{d G(q^2)}{d q^2} \right |_{q^2=0}  \approx \frac{21.5}{(\tilde{\alpha}_s M_Q)^2}   \\
\langle r^4 \rangle_{\rm baryon}  &= 60 \left . \frac{d^2 G(q^2)}{d (q^2)^2} \right |_{q^2=0} \approx \frac{890.}{(\tilde{\alpha}_s M_Q)^4}
\end{split}
\end{equation}
where the numerical values were obtained by  the mean-field form factor using the wave function in Eq.~(\ref{approxwf}) and thus are accurate at leading order in the double expansion in $1/N_c$ and $\Lambda_{\rm QCD}/M_Q$.

\begin{figure}
\includegraphics[width=3in]{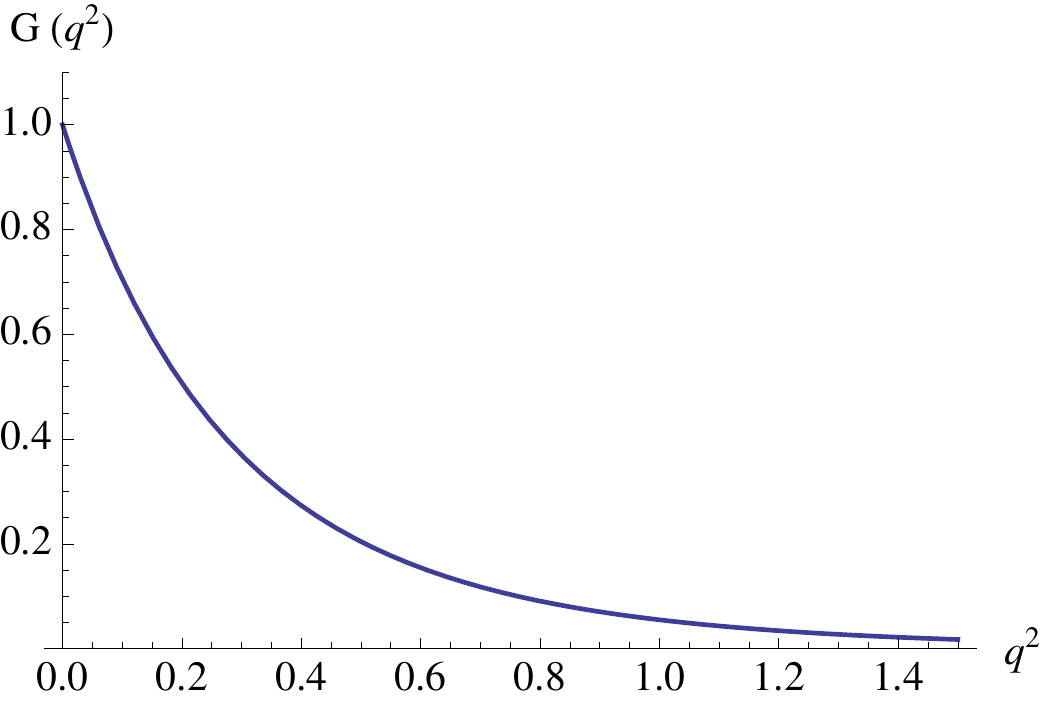}
\caption{The baryon form factor in the large $N_c$ and heavy quark limits.  In the plot, $q^2$ is given in units of   $(\tilde{\alpha}_s M_q )^{2}$  }
\label{fig:fig2}
\end{figure}

\subsection{Degeneracy}

Due to the spin and flavor degrees of freedom there are a large number of distinct species of baryons and, in  this limit, the masses of many of these  degenerate.  Recall that baryon mass is insensitive to either the spin or the flavor of the quarks making up the baryon.  The reason for this is clear---by hypothesis the mass of the quarks are degenerate and in the heavy quark limit the spin of the quarks decouple from the other degrees of freedom in the problem.  Moreover as discussed above in the mean-field state the spatial wave functions are all identical s-waves.  Combining this with the anti-symmetry from Fermi statistics and the anti-symmetry associated with the color degree of freedom implies that, at leading order in the double expansion, the degeneracy of the ground state baryon is given by the number of distinct symmetric  spin and isospin   configurations that can be constructed from $N_c$ quarks.

There are $2 N_f$ possible spin-isospin states for each quark.  The symmetry under exchange implies that each state in the degenerate multiplet is fully specified by the number of quarks in the baryon, with each of the $2N_f$ possibilities subject to the constraint that the total number of quarks is $N_c$.  Thus the degeneracy is given by the following bionomial coefficient
\begin{equation}
d=\left (   \begin{array}{c} N_c + 2 N_f \\ 2 N_f  \end{array} \right )  \; .\label{degen }\end{equation}

Of course, this degeneracy is split.  Firstly if the quark  masses are not precisely degenerate there is a splitting due to the mass differences.  However, even in the exact flavor symmetric limit there is a splitting due to color hyperfine splittings.  Characteristically the maximum splitting of the full multiplet is characteristically given by $ \Delta M_{\rm max} \sim \tilde{\alpha}_s N_c \lambda_{\rm QCD}/M_Q$.  Note that in this formalism the heavy quark limit is taken prior to the large $N_c$ limit and thus  the total splitting remains small in the combined limit.

\subsection{The value of $\tilde{\alpha}_s$}

The calculations detailed above treat $\alpha_s$ as a constant.  Of course, in reality there is a scale anomaly.   Thus  the coupling constant  runs, albeit logarithmically.     Using standard renormalization group analysis one sees that at one loop and at large $N_c$
\begin{equation}
\tilde{\alpha}_s = \frac{12 \pi}{11 \rm{Log}\left (\frac{Q^2}{\Lambda_{\rm QCD}^2} \right ) }
\end{equation}
where $Q$ is the momentum scale of the process and $ \Lambda_{\rm QCD}$ is the QCD scale.  Thus, in the formalism used above,
 $ \tilde{\alpha}_s$ is to be interpreted as being evaluated at the momentum scale of the problem.

Of course, the question of what constitutes the optimal choice of scale is not completely well posed.  The effective value of the coupling differs for different parts of the wave function.   However, over the region of the wave function  which makes the dominant contributions,  $\tilde{\alpha}_s$ changes by only a small amount since the running  is logarithmic and the inverse length scale of the wave function is much smaller than $\Lambda_QCD$.  Thus, the ambiguity is of relatively small importance provided that the $\tilde{\alpha}_s M_Q$ is high enough.  The inclusion of higher order corrections in $\tilde{\alpha}_s$ would further reduce the ambiguity.

The relevant momentum scale of this problem is $\tilde{\alpha}_s M_Q$, which is the inverse of the characteristic length scale of the wave function.  This yields a transcendental equation for $\tilde{\alpha}_s$:
\begin{equation}
\label{trans}
\tilde{\alpha}_s = \frac{12 \pi}{11 \rm{Log}\left (\tilde{\alpha}_s^2  \frac{M_Q^2}{\Lambda_{\rm QCD}^2} \right ) } \; .
\end{equation}
 The analysis in this paper is based on  the regime in which $ \frac{M_Q}{\Lambda_{\rm QCD}} \gg 1 $ and  $\tilde{\alpha}_s \ll 1$.   These conditions might be satisfied to reasonably good approximation even well  away from  the extreme large mass limit in which $ \rm{Log}\left (  \frac{M_Q^2}{\Lambda_{\rm QCD}^2} \right ) \gg 1$;  in those cases it is necessary to solve Eq.~(\ref{trans}) numerically in order to find $\tilde{\alpha}_s$, given the value of $M_Q$ .   Using this value of $\tilde{\alpha}_s $ in Eq.~(\ref{baryon}) yields the baryon mass.

However, in the regime  where $ \rm{Log}\left (  \frac{M_Q^2}{\Lambda_{\rm QCD}^2} \right ) \gg 1$ , it is possible to  estimate analytically $\tilde{\alpha}_s$.   The key point is that  in this regime,  $ \rm{Log}\left (\tilde{\alpha}_s^2  \right )/  \rm{Log} \left (\ \frac{M_Q^2}{\Lambda_{\rm QCD}^2} \right ) \ll 1$; thus we can rewrite
 Eq.~(\ref{trans}) as
\begin{equation}
 \label{trans2}
 \begin{split}
&\tilde{\alpha}_s = \frac{12 \pi}{ 11  \left  (\rm{Log}\left (\ \frac{M_Q^2}{\Lambda_{\rm QCD}^2} \right ) + \rm{Log}\left (\tilde{\alpha}_s^2  \right ) \right )  } \\
&\approx \frac{12 \pi}{ 11  \rm{Log}\left (\ \frac{M_Q^2}{\Lambda_{\rm QCD}^2} \right )   } \left ( 1 - \frac{\rm{Log}\left (\tilde{\alpha}_s^2  \right )}{  \rm{Log}\left (\ \frac{M_Q^2}{\Lambda_{\rm QCD}^2} \right ) }    \right) .
\end{split}
\end{equation}
where we have dropped terms of relative order  $ \rm{Log}^2\left (\tilde{\alpha}_s^2  \right )/  \rm{Log}^2\left (\ \frac{M_Q^2}{\Lambda_{\rm QCD}^2} \right ) $.   Using the leading order result to estimate the value  $ \tilde{\alpha}_s^2 $ in the subleading term yields
 \begin{equation}
 \label{trans3}
 \begin{split}
\tilde{\alpha}_s  & \approx \frac{12 \pi}{ 11  \rm{Log}\left (\ \frac{M_Q^2}{\Lambda_{\rm QCD}^2} \right )   }\\
& -\left(  \frac{12 \pi}{11}  \right) \frac{     \left (     \rm{Log}\left ( \frac{12 \pi}{ 11   }  \right )      -  \rm{Log}\left ( \rm{Log}\left (\ \frac{M_Q^2}{\Lambda_{\rm QCD}^2} \right )    \right )     \right )    }{   \rm{Log}^2\left (\ \frac{M_Q^2}{\Lambda_{\rm QCD}^2} \right )   }  \, .
\end{split}
\end{equation}
It is apparent that the second term in Eq.~(\ref{trans3}) is indeed suppressed compared to the leading one in the regime  $ \rm{Log}\left (  \frac{M_Q^2}{\Lambda_{\rm QCD}^2} \right ) \gg 1$.   In this regime one can use Eq.~(\ref{trans3}) to find   $\tilde{\alpha}_s $ which, in combination with  Eq.~(\ref{baryon}), yields the baryon mass.

 \section{Baryonic matter in the combined large $N_c$ and heavy quark limits}

Baryonic matter is  the analog of infinite nuclear matter in this artificial world with many colors and heavy quarks.  In describing baryonic matter in this regime it is necessary to develop a formalism which correctly encodes the leading orders in both the large $N_c$ and heavy quark expansions.  In ref.~\cite{Witten:1979kh} Witten  assumed implicitly that the argument justifying mean-field theory for the single-nucleon case also holds for the many-nucleon problem.  In particular, ref.~\cite{Witten:1979kh} assumed that the appropriate mean-field wave function for a two-baryon scattering-state is a Slater determinant  composed of two distinct clusters, each composed of $N_c$ quarks in identical spatial wave functions coupled to a color singlet.     Before proceeding, it is probably worth remarking that although this is quite plausible, it is not trivially obvious.  The multi-baryon problem  is more complicated  than the analogous single-baryon case  since the bosonic substitution made in Subsec. \ref{bosonic} no longer holds.   However, the qualitative reasoning underlying this assumption is quite plausible and  we will assume it  to be correct for the remainder of this paper.  In particular, it  will be assumed that, as with the single-nucleon case, such mean-field calculations will accurately give the mass (up to corrections of relative order $1/N_c$) while not accurately describing the wave function.

In the analysis of the single-baryon problem,  the spin-flavor structure of the baryon was irrelevant---provided that the state was symmetric.  The heavy-quark limit leads to a decoupling of the spin degree of freedom while the degeneracy of the quark masses leads to a flavor symmetry.  Thus the only possible effect of the spin-flavor symmetry is due to the effect of the Pauli principle.  In the single-baryon case, anti-symmetry of the wave function with respect to color means that the space-spin-flavor part of the wave function is symmetric.  Since the mean-field treatment necessarily leads to a symmetric spatial wave function, the spin-flavor state is automatically symmetric.    However, in a certain strict sense, mean-field theory of the Hartree-Fock type  applies only 
to  a subset of such baryons: namely states in which  all of the quarks are an identical spin-flavor state. These could be  states where each quark is in well-defined projections in spin and flavor space separately or  one in which spin and flavor are correlated as, for example, in a hedgehog.     In other spin-flavor configurations the state cannot be described as a single Slater determinant since there are correlations between spin, flavor and color.  Thus, one should envision the myriad of states in the single-baryon system as a two-part problem: first, one uses a standard Hartree-Fock analysis based on a single Slater determinant to compute the energy.  One then subsequently  can make transformations on top of the single Slater determinant state which do not alter the energy (for heavy quarks) to consider states with  spin-flavor correlations.

In this section, we study the energy density of cold baryonic matter using mean-field techniques, which as noted above, are assumed to hold in the combined large $N_c$ and heavy quark limits.  Following Witten, we will assume that the state can be written as a single Slater determinant and that any effects of spin-flavor correlations can by obtained by acting on top of a single Slater determinant without changing its energy.  In general, even with this assumption, the problem is rather complicated.   However, it greatly simplifies at low density.  We show that there exists a phase which is repulsive and has an interaction energy per baryon which is exponentially small at low densities.  This phase is shown to be at least metastable.  This phase can be the true ground state depending on whether baryonic matter in this limit saturates.  While we leave this question to future research, we will briefly consider the effects of $1/N_c$ corrections and show that a phase in which baryonic matter saturates is likely to emerge at next-to-leading order in $1/N_c$ but that density and binding energy per baryon of saturated matter both go to zero as $N_c$ goes to infinity.

\subsection{A simplified problem}

 As noted above,  for the single-baryon case the only way to construct a single Slater determinant while imposing a color singlet structure is to have all the quarks in an identical spin-flavor state.  The problem under consideration in this subsection is to determine  what happens when there is a non-zero density of identical baryons in this class and no other baryons in the system.  To state the problem more precisely, we seek the ground state  of a system with a fixed density of quarks each with the same spin-flavor state.  We will turn to the more general problem without this restriction later.  It is worth noting that the present  problem is not well posed in QCD with finite quark masses since the spin quantum number is not conserved.  However, in the limit where $M_Q \rightarrow \infty$, the spin degree of freedom decouples and spin becomes an emergent symmetry with an associated conservation law; in this limit the problem is well posed.

By fixing each quark to have an identical spin-flavor state---even quarks associated with different baryons---the problem simplifies considerably:  the spatial wave function for two quarks in the systems with identical color needs to be anti-symmetric. Moreover, at the mean-field level,  by construction, there are two and only two distinct wave functions for the two  quarks of the same color; anti-symmetry then implies that they are orthogonal.  Thus for a system with a total baryon number, $N_{\rm bary}, $ greater than unity we can consider a color singlet state of the following form:
  \begin{equation}
  |\Phi \rangle =\prod_{j=1}^{N_{\rm bary} }\prod_{a=1}^{N_c} \left ( \int {\rm d}^3r \, \psi_j(\vec{r}) \hat{\psi}_{\rm sf; a}^\dagger(\vec{r})        \right )|{\rm vac} \rangle
  \end{equation}
  where $a$ indicates color, the subscript sf indicates a particular spin-flavor state and $j$ distinguishes between the various single particle states; the wave-functions $\psi_j$ are taken to be orthonormal: $\int {\rm d}^3r \, \psi_j^*(\vec{r}) \psi_k(\vec{r}) = \delta_{jk}$.
  To make the infinite matter problem concrete, one should envision keeping the number of baryons, $N_{\rm bary}$, large and finite and then imposing boundary conditions restricting all of the wave functions to have support only within some volume, $V$.  The limit, $V\rightarrow \infty, \, \,  \, N_{\rm bary} \rightarrow \infty$ with $N_{\rm bary}/V$ held fixed is then taken at the end of the problem.  The many-body state, $|\Phi \rangle$, can then be inserted to into the nonrelativistic reduction of the QCD Hamiltonian yielding the following energy functional:
  \begin{widetext}
  \begin{equation}
  \label{manybaryon}
	\begin{split}
		\langle \psi | \hat{H} | \psi \rangle &= {\cal E} V =  N_c \, N_{\rm bary}\,  M_Q \,  +
N_c \sum_{j=1}^{N_{\rm bary}}   \left ( \int{d^3r
\frac{|\nabla \psi_j|^2}{2M_Q}} - \frac{1}{2}  \tilde{\alpha}_s
\int{d^3 r \, d^3 r' \frac{|\psi_j(\vec{r})|^2 | \psi_j (\vec{r'})|^2}{|\vec{r'} - \vec{r'}|}}   \right )  \\
 - \frac{1}{2}   N_c \tilde{\alpha}_s  & \sum_{j=1}^{N_{\rm bary}}  \sum_{  k=1}^{N_{\rm bary}}   \, \left ( \delta_{ij}-1  \right)
\int{d^3 r \, d^3 r' \frac{\psi_j^*(\vec{r})  \psi_j(\vec{r'}) \psi_k^*(\vec{r'})  \psi_k(\vec{r})}   {|\vec{r'} - \vec{r'}|}}
\end{split}
	\end{equation}
	  \end{widetext}
subject to the constraint  $\int {\rm d}^3r \, \psi_j^*(\vec{r}) \psi_k(\vec{r}) = \delta_{jk}$.  The first term represents the dominant energy arising from the mass of the heavy quarks.  The second term may be interpreted as the interaction energy of quarks within one baryon summed over all baryons and the third term as the interaction energy between baryons.  Note, however,  in general the second term differs from the interaction energy between quarks in an isolated baryon since the quark wave functions rearrange themselves due to the presence of the additional baryons.

The next step is to find the minimum energy configuration subject to the constraints.  It is generally believed that in the large $N_c$ limit baryonic matter will form a crystal.  The underlying reason for this is clear if one thinks about baryonic matter using a simple picture of baryons interacting via potentials: baryons are heavy and their interactions strong.  In such a regime one generically expects baryons to find the minimum of the potential created by the others and sit there.  The kinetic energy of the baryon---which is what typically fights against crystallization---scales as $1/N_c$  and is thus suppressed.   At a more fundamental level the large $N_c$  limit is, in an important sense, classical:  the expectation value of a product of color singlet operators is the product of the expectation values of the operators.  Thus, quantum correlations of the physical degrees of freedom are suppressed.  However, spatial correlations are not; expectation values of operators can vary in space.  To minimize energy, the system will generally exploit its freedom to build in spatial variations yielding either crystals or amorphous solids.  For a wide class of systems, crystals  have lower energy than amorphous solids.  The general behavior  of baryonic matter forming a crystal is seen explicitly in Skyrme-type models \cite{SC} which are designed to reproduce the large $N_c$ scaling behavior of QCD.   Moreover, the heavy quark limit also leads to heavy baryons and should not affect the qualitative result that crystalline phase is expected.  Indeed the heavy quark limit should act to reinforce this expectation.

To compute the energy density, we will need an explicit ansatz for the crystalline form.   If a  low density phase of baryonic matter exists,  the appropriate ansatz which minimizes the energy is easy to find.  As seen from Sec. \ref{1baryon} the wave function for a single baryon drops off exponentially at long distance.  Moreover  Eq.~(\ref{manybaryon}) implies that the interaction energy between baryons depends on the product of the two wave functions at the same spatial point.  If the wave functions of interacting baryons in a low density system are qualitatively similar to free baryons, one expects that the interaction energy between two baryons in the system to drop exponentially with the distance between baryons.  Effects of the Pauli principle will also be exponentially small.   Thus, in a regular crystalline configuration at low densities all interactions are expected to be exponentially small.  Moreover, nearest-neighbor interactions although exponentially small are also exponentially larger than next-to-nearest-neighbor interactions.   There are two possibilities to consider: the case where the nearest neighbor interactions are attractive and the case where they are repulsive.  For the attractive case, the low density matter is not stable and the system will collapse; low density baryonic matter does not exist.  However, if the interaction is repulsive---as we shall show below it is---then low density baryonic matter exists at least as a metastable phase provided that there is an external pressure to prevent the system from expanding.  Moreover, the crystal structure which minimizes the energy density for fixed baryon density is the one which maximizes the nearest-neighbor distance at fixed density.

Let us denote the nearest  neighbor distance as $d$.   By dimensional analysis, it is related to the baryon density by
$   \rho=\frac{c}{d^3}$
    where $c$ is a constant that depends on the crystal structure.   It is conventional in crystallography to describe structures in  terms of their atomic packing fraction $P$ which gives the maximum fraction of  the volume filled by rigid non-overlapping spheres of fixed radius arranged in the given crystalline configuration.  It is related to the constant $c$ by $c=\frac{3P}{4\pi}$.  Thus,
    \begin{equation}
    \label{apf}
    d= \left ( \frac {4 \pi }{3 P \rho} \right)^{1/3} \; ,
    \end{equation}
 and the configuration with the largest  nearest-neighbor spacing  for fixed baryon density is the one with the largest value for the atomic packing-factor, $P$.   It has been known since Gauss that the maximum value for  $P$ in any crystalline configuration is  $P=\frac{\pi}{\sqrt{18}}$ which occurs for both hexagonal close-packed (HCP) lattices and for face-centered cubic (FCC)  lattices.  Thus, provided that a stable low density phase exists,  the nearest-neighbor distance will be given by
    \begin{equation}
    \label{d}
    d= \left ( \frac {4 \sqrt{2}}{ \rho} \right)^{1/3} \; .
    \end{equation}

 The preceding analysis is based on the assumption  that the system forms a crystal. As noted previously, the arguments based on large $N_c$ imply that the system will either form a crystal or an amorphous solid.  The argument given above can serve to rule out the possibility of an amorphous solid.  Note that the structure with the lowest energy is the one with the maximum atom packing fraction, $P$.  Since Kepler first conjectured it in the seventeenth century, it has been believed  that  maximal close-packing occurs for the HCP and FCC crystalline structures and not for an aperiodic configuration.    A recent proof that this is indeed the case has been constructed, albeit  one requiring a computer to verify a large number of cases\cite{Kepler}.

 To proceed, we need to compute the interaction energy per baryon or, equivalently, the interaction energy density divided by the baryon density ${\mathcal E}_{\rm int}/\rho$ .  As noted above, at low densities, the interaction energy between baryons is entirely dominated by nearest neighbor interactions.  Thus, the energy per baryon from the mean-field energy functional Eq.~ (\ref{manybaryon}) reduces to
 \begin{widetext}
    \begin{equation}
    \label{eq:bigeq}
    \begin{split}
    \frac{{\mathcal E}_{\rm int}}{\rho} =& N_c\int{d^3 r \; \frac{\nabla \psi^*(\vec{r})\nabla \psi(\vec{r})}{2 M_Q}}
    -\frac{N_c}{2}  \, \tilde{\alpha}_s \, \iint d^3 r\;d^3 r' \;\frac{\psi^*(\vec{r})\psi(\vec{r})\psi^*(\vec{r'})\psi(\vec{r'})}{|\vec{r}-\vec{r'}|}\\
    &-\frac{N_c}{2}  \, \tilde{\alpha}_s \,\sum_i\iint d^3 r \;d^3 r' \;\frac{\psi^*(\vec{r})\phi_i^*(\vec{r'})
    \psi(\vec{r'})\phi_i(\vec{r})+\psi^*(\vec{r'})\phi_i^*(\vec{r})\psi(\vec{r})\phi_i(\vec{r'})}{|\vec{r}-\vec{r'}|}
    \end{split}
    \end{equation}
    \end{widetext}
   $\psi$ is the wave function of the quarks in one of the baryons,  the summation is over nearest neighbors  and $\phi_i$ represents  the wave function of the quarks in the nearest neighbor.  We wish to minimize this, subject to constraints that the wave functions are orthonormal and are centered a distance $d$ apart.  As a first step let us  find the variational equation which minimizes ${\cal E}/{\rho}$ subject to the constraint of orthonormality but without yet imposing the condition that the wave functions are centered a distance $d$ apart.  A trivial application of the calculus of variations on ${\cal E}/{\rho}$ with respect to $\psi^*$ yields
    \begin{equation}
    \begin{split}
    \label{eq:variedE}
  &  \left[-\frac{\nabla^2}{2 M_Q}- \, \tilde{\alpha}_s \, \int d^3 r' \;\frac{\psi^*(\vec{r'})\psi(\vec{r'})}{|\vec{r}-\vec{r'}|}-\epsilon\right]\psi(\vec{r})= \\
& \sum_i
 \left (   \Lambda_i  \phi_i(\vec{r}) +  \, \tilde{\alpha}_s \, \int d^3 r' \;\frac{\phi_i^*(\vec{r'})\psi(\vec{r'})}{|\vec{r}-\vec{r'}|  } \phi_i(\vec{r}) \right ).
   \end{split}
    \end{equation}
where $\epsilon$ is a Lagrange multiplier enforcing the normalization of $\psi$, and the $\Lambda_i$ are Lagrangian multipliers enforcing the orthogonality of the wave function of the baryon, $\psi$, and its nearest neighbors, $\phi_i$.

Note that the terms on the left-hand side of Eq.~(\ref{eq:variedE}) would be zero for a wave function that minimizes the single-baryon problem as in Eq. (\ref{eq:diffeq}).   Thus it is natural to treat the right-hand side of Eq.~(\ref{eq:variedE}) as a perturbation.   Treating this perturbatively has the virtue of keeping the wave functions well localized a distance $d$ apart.   There are two effects in this perturbation: one arising from the Pauli principle through the orthogonality constraint encoded by the Lagrangian multipliers, and the other from the explicit interaction seen in the integral term.  It is highly plausible that the  second effect is parametrically smaller than the first by a factor which scales as $(\alpha_s M_q d)^{-1}$:  the integral depends on $|\vec{r}-\vec{r'}|^{-1}$ for wave functions which peak a distance $d$ apart and which have a natural distance scale $\sim (\alpha_s M_q )^{-1}$ . Accordingly, we will drop the term at this stage and subsequently verify that the effect is indeed small for large $d$ ({\it i.e.}, small $\rho$)

The Lagrange multipliers $\Lambda_i$ have dimensions of mass.  It is convenient to rewrite them as
\begin{equation}
\Lambda_i =\epsilon_0 \lambda_i
\end{equation}
where $\epsilon_0 \equiv  -\frac{3 \tilde{\alpha}_s^2 M_Q }{4} \frac{{\cal V}^2}{\cal T}  \approx   .16277, \; \tilde{\alpha}_s^2 M_Q$ is the eigenvalue of  the unperturbed single-baryon problem of Eq.~(\ref{eq:diffeq}) and $\lambda_i$ is dimensionless.  Since the overlaps between the wave functions are expected to be exponentially small, the  $\lambda_i$ are useful as expansion parameters.  Thus, we can write
    \begin{equation}
        \label{eq:expansions}
    \begin{split}
    \psi&=\psi^{0}+\sum_j \lambda_j \psi_j^{1}+\mathcal{O}(\lambda^2)\\
        \phi_j&= \phi_j^{0}+\lambda_j \phi_j^{1}+\mathcal{O}(\lambda^2)\\
    \epsilon&=\epsilon^{0}+\sum_j \lambda_j \epsilon_j^{1}+\mathcal{O}(\lambda^2)
    \end{split}
    \end{equation}
  where the sum over $j$ indicates nearest neighbors  and the shifts in the $\phi_j$ are only those induced by its interaction with $\psi$.  Note that at lowest order, the wave functions are simply those of noninteracting baryons so that  $\phi_j^0(\vec{r})=\psi^0(\vec{r}-d \hat{n}_i)$ where $\hat{n}_i$ is a unit vector in the direction between the reference particle and its nearest neighbor and $d$ is the distance between nearest neighbors and $\psi^0$ is well approximated by Eq.~(\ref{approxwf}).  It is worth noting that one can only use this expansion  consistently up to first order in the $\lambda_j$; higher order term will turn out to be parametrically  smaller than the effects of next-to-nearest neighbor interactions which we are neglecting here.  Inserting the expansion of Eqs.~(\ref{eq:expansions}) into Eq.~(\ref{eq:variedE}), neglecting effects which are parametrically suppressed by factors of $(\tilde{\alpha}_a M_q d)^{-1}$ and solving for the interaction energy to first order in $\lambda_j$, one obtains
    \begin{equation}
    \label{eq:deltaE}
    \delta E \equiv \sum_j \lambda_j  \epsilon_j^1
= \sum_j\frac{|\langle\psi^0|\phi_j^0\rangle|^2}{2 \langle\psi^0|\frac{1}{\frac{\hat{p}^2}{2 M_Q}
    -\epsilon^0}|\psi^0\rangle}\\
    \end{equation}
where the wave functions are represented in Dirac notation and $\hat{p}$ is the momentum operator.
The derivation of this equation is straightforward but somewhat involved, and can be found in Appendix A.

 The numerators in Eq.~(\ref{eq:deltaE}) are the squares of the relevant  overlap integrals $\mathcal{A}_j\equiv\langle\psi^0|\phi_j^0\rangle= \int d^3r \;\psi^0(\vec{r})\;\psi^0(\vec{r}- d \hat{n}_j )$.
 Thus, $\mathcal{A}_j$ is a measure of orthogonality of two wave functions $\psi^0$ and $\phi_j$ and is exponentially small.   Since the original wave functions are spherically symmetric, $\mathcal{A}_j$ depends only on $d$ and not on $\hat{n}_j$---the direction connecting the nearest neighbors.  Thus $\mathcal{A}_j \equiv \mathcal{A}$ for all $j$ and all terms in Eq.~(\ref{eq:deltaE}) are equal.  Since $\mathcal{A}$ does not depend on the direction, without loss of generality we can take the direction to be along the z axis and  exploit cylindrical symmetry:\begin{widetext}
    \begin{equation}
    \label{eq:overlapintegral}
    \mathcal{A}
   ={2\pi}\int_{-\infty}^\infty\int_0^\infty{dz\, dr\,
    r\,\psi^0(\sqrt{r^2+z^2})\psi^0(\sqrt{r^2+(d-z)^2})}.
    \end{equation}
    \end{widetext}
    Moreover, it is easy to see that on the scale of $d$ the integrand for the $r$ is sharply peaked and can be evaluated using the standard logic underlying the steepest descents approximation; the relative error in making this approximation is of order $(\tilde{\alpha}_s M_Q d)^{-1}$.   The $z$ integral  can then be evaluated straightforwardly  yielding,
    \begin{equation}
    \begin{split}
    \label{eq:overlapsol}
    \mathcal{A}(\rho)=\gamma\tilde{ \rho} ^{-\frac{\sqrt{2}}{3 \sqrt{-\epsilon^0 }}}\exp \left (2(\frac{\pi}{3})^{2/3}\sqrt{-\epsilon^0}\tilde{\rho}^{-1/3 } \right )\\
    {\rm with}\; \tilde{\rho} \equiv \frac{\rho}{(\tilde{\alpha}_s M_Q)^3}    \; \; {\rm and} \;  \gamma\approx0.044984 \; ;
    \end{split}
    \end{equation}
 we have used Eq.~(\ref{d}) to eliminate $d$ in favor of  $\rho$.
 
    The numerators in Eq.~(\ref{eq:deltaE}) are given by $|\mathcal{A}|^2$.
    The denominators in  Eq.~(\ref{eq:deltaE}),  $ \mathcal{B}\equiv 2\langle\psi_s|\frac{1}{\frac{\hat{p}^2}{2 M_Q}
    -\epsilon^0}|\psi_s\rangle$ , are easily evaluated:
\begin{equation}
\begin{split}
\mathcal{B} &=
   2   \int \frac{d^3p}{(2\pi)^3}\;\frac{|\langle p|\psi_s\rangle|^2}{\frac{\hat{p}^2}{2 M_Q}-\epsilon^0}\\
 {\rm with} \; &   \langle p|\psi_s\rangle=\int d^3\vec{x}\;e^{-i \vec{p}\cdot \vec{x}}\psi_s(\vec{x})
    \end{split}
    \end{equation}
  yielding  $\mathcal{B}\approx 76.914 (\tilde{\alpha}_s^2 M_Q)^{-1}$

 Inserting $\mathcal{A}$ and $\mathcal{B}$ into Eq.~(\ref{eq:deltaE}) and using the fact that all four nearest neighbor interactions are equal yields an interaction energy per baryon of
\begin{widetext}
\begin{equation}
    \label{eq:finalresult}
    \delta E \equiv \frac{\mathcal{ E}_{\rm int} }{\rho} \approx    .00042858\,  N_c \, M_Q\,  \tilde{\alpha_s}^2  \,
\tilde{\rho}^{2.3369}\exp(-2.0332\tilde{\rho}^{1/3})\;
\;  \; \; {\rm with}\; \tilde{\rho} \equiv \frac{\rho}{(\tilde{\alpha}_s M_Q)^3}.
    \end{equation}

Equation (\ref{eq:finalresult}) gives the interaction energy per baryon  for this simplified problem.  In its derivation we neglected the explicit potential energy contribution to the perturbation in Eq.~(\ref{eq:variedE}) and only included the effect of the Pauli principle.  Let us now show {\it a posteriori} that this is justified.  The energy associated with this potential energy, at first order in the perturbation expansion, is given by
\begin{equation}
V_{\rm int}= -4 {N_c}  \, \tilde{\alpha}_s \int d^3 r \;d^3 r' \;\frac{\psi^0(\vec{r})\psi^0(\vec{r'}-\hat{n} d)
    \psi^0(\vec{r'})\psi^0(\vec{r}-\hat{n} d)}{|\vec{r}-\vec{r'}|}
\end{equation}
\end{widetext}
where $\hat{n}$ is an arbitrary unit vector and we have used the fact that the wave functions are real.  Note that up to a sign, the form of this is precisely that of an ordinary electrostatics problem for the repulsive Coulomb energy of a charge distribution proportional to $\psi_0(\vec{r})\psi_0(\vec{r}-\hat{n} d)$; the charge  distribution is cylindrically symmetric and has a characteristic length of $d$ and a characteristic width of order $(\tilde{\alpha}_s M_Q)^{-1}$.  Note that the total ``charge'' in this electrostatic problem is $\mathcal{A}$.  It is  straightforward to show that for $\tilde{\alpha}_s M_Q d  \gg 1$ that
$V_{\rm int} \sim |\mathcal{A}|^2  \log ( \tilde{\alpha}_s M_Q d   ) d^{-1}$ (up to overall constants).  Comparing this to the expression for  $\delta E$, one sees that
\begin{equation}
\frac{V_{\rm int}}{\delta E} \sim  \frac{\log ( \tilde{\alpha}_s M_Q d   ) }{  \tilde{\alpha}_s^2 M_Q d} \; ,
\end{equation}
which goes to zero at large $d$.

It is not immediately clear whether the phase described in this section is absolutely stable.  At the  large interparticle distances studied here, the repulsive interactions induced by  the Pauli principle at the quark level dominate over the color-Coulomb attractive interactions from the Fock term.  However, as the densities increase so does the effect of the color-Coulomb interaction.  When the densities reach the order of $( \tilde{\alpha}_s M_Q)^3$ , both the repulsive effects induced by the Pauli principle at the quark level and the attractive Fock interactions contribute to the interaction energy per baryon an amount which is parametrically of order $ \tilde{\alpha}_s M_Q^2$.  Without detailed calculations it is impossible to know which effect is larger, {\it i.e.} whether the net  interaction energy is positive or negative.  Note that such  calculations are highly nontrivial in this regime since the system need not be dominated by nearest neighbor interactions.  If it happens that interaction energy everywhere in this regime is repulsive, then the phase we have computed  is presumably  absolutely stable.  If, however,  the interaction energy is attractive somewhere  in this regime, then the phase we have computed is not absolutely stable; the system can lower energy by collapsing to these higher densities.    However, even if it turns out that the low density phase is not absolutely stable in a global sense, the phase is {\it metastable}; all  local changes in the system will necessarily raise the energy per baryon.

\subsection{The full problem}

In the last subsection we found the energy density for a system in which all of the quarks were constrained to be in the same spin-isospin state.  In this section we relax this constraint to find the energy density of a system with fixed (low) total baryon density.  Again, we will assume the system is in the form of a single Slater determinant.   To create localized color singlet clusters (baryons) within a Slater determinant we still need to consider products over color of single-particle states with a single spin-space-flavor state and then take a product of these.  The difference in this problem is that space-flavor combinations in these clusters need not be the same for all clusters.  Thus our state is of the form
  \begin{equation}
  |\Phi \rangle =\prod_{j=1}^{N_{\rm bary} }\prod_{a=1}^{N_c} \left ( \int {\rm d}^3r \, \psi_j(\vec{r}) \hat{\psi}_{{\rm sf}_j; a}^\dagger(\vec{r})        \right )|{\rm vac} \rangle
  \end{equation}
where ${\rm sf}_j$ represents the spin-flavor state of the quarks in baryon $ j$ and $a$ is color.

Now let us consider the interaction between two baryons.   At low densities the dominant interaction is the Pauli principle at the quark level.  Note that if the spin-flavor state of the quarks in the two baryons are orthogonal to each other the Pauli principle does not apply and there is no Pauli repulsion between them.  If they are not orthogonal, there is repulsion due to the Pauli principle.    Thus we can decrease the energy per baryon relative to the simplified problem of the previous subsection by simply constructing $2 N_f$ copies of the state  from the simplified problem, each associated with an orthogonal spin-flavor state for the quarks:
  \begin{equation}
\label{ansatz}
  |\Phi \rangle =\prod_{j=1}^{N_{\rm bary} }\prod_{a=1}^{N_c} \prod_{s=1}^2 \prod_{f=1}^{N_f} \left ( \int {\rm d}^3r \, \psi_j(\vec{r}) \hat{\psi}_{s,f,a}^\dagger(\vec{r})        \right )|{\rm vac} \rangle \; .
  \end{equation}
Note that by construction this state is  both a spin and flavor singlet since for each spatial wave function and each color one has a fully  antisymmetric state in flavor and spin separately.
The energy density is trivial to compute.  As noted above there is no Pauli repulsion between baryons with orthogonal spin-flavor states of the quarks.  It is also trivial to see that the color-Coulomb Fock interaction vanishes  between baryons  with the same spatial wave functions but with  orthogonal spin-flavor states of the quarks.  Thus, the total  interaction energy is simply $2 N_f$ times the interaction energy of the simplified problem for a simplified problem with a density of $\rho/(2 N_f)$.  The interaction energy per baryon  is then given by
\begin{equation}
    \label{eq:finalresultfull}
\begin{split}
     \frac{\mathcal{ E}_{\rm int} }{\rho} &  \approx    .00042858\,  N_c \, M_Q \,  \tilde{\alpha_s}^2  \,
\tilde{\rho}^{2.3369}\exp\left ( -2.0332 \tilde{\rho}^{1/3} \right)\\
  &  {\rm with}\; \tilde{\rho} \equiv \frac{\rho}{2 N_f (\tilde{\alpha}_s M_Q)^3}.
    \end{split}
    \end{equation}
Note that at low densities this phase is much smaller than for the simplified problem.  As with the simplified problem we do not know whether the low density phase described in this section  is absolutely stable or merely  metastable.  However, we do know that it is at least metastable.

In concluding that this phase is at least metastable an issue arises which does not come up in the simplified problem:   the role of the spin-flavor degree of freedom.  Note that the ansatz of Eq.~(\ref{ansatz}) has baryons which are not well separated.  Of course, the baryons which are not well separated are also not interacting since they are made of quarks which are orthogonal in spin-flavor.  However, to conclude that the system is metastable, we need to know that {\it all}  local variations in the state consistent with the Slater determinant form (which by assumption we take to yield the  energy at leading order in the $1/N_c$ expansion) which keep the average baryon density fixed act to raise the energy.  Clearly, in verifying this we do not  need to consider interactions between well-separated baryons.  These act as they do in the simplified problem.   However, one does need to consider what happens if any of the $2 N_f$  baryons with distinct spin-flavor states which are localized in the same spatial state are varied so that they are no longer completely orthogonal in spin-flavor space.   Of course, if the system does this it must  simultaneously alter its spatial configuration to respect the Pauli principle at the quark level.  Clearly,  the effect due to the  Pauli principle is entirely repulsive.  When the system makes local variations  at lowest order, this Pauli effect is the only effect contributing.  Larger scale variations might in principle yield attractive effects which offset the Pauli repulsion and lead to a net lower energy per particle than the phase considered here.  Thus, we do not know the system is globally stable.  However, since all small variations only involve the repulsive interactions, the system is at least metastable.

\subsection{Subleading contributions and saturation\label{SL}}

So far in this section,  we have shown that there exists a  low-density phase of QCD  in the combined large $N_c$ and $1/N_c$ limits for which the interactions are repulsive and thus a low-density phase exists which is at least metastable.  Such a phase clearly does not lead to saturating matter in the manner of QCD  at $N_c=3$ and physical quark masses in which the matter is self-bound and  stable with nonzero density and zero external pressure.  We do not know at this stage whether the phase we have studied is absolutely stable in the combined limit or whether there are attractive interactions at shorter distances that might give rise to a saturating phase.

 Remarkably, however, it is possible to conclude that as the combined  large $N_c$  and heavy quark limit is approached,  the system must have a saturating phase provided  that the lightest glueball in the spectrum at large $N_c$ is scalar.

At first blush, it may seem that, by construction, subleading effects in $1/N_c$ must have a small effect on the interaction energy and thus as the limit is approached they are extremely unlikely to affect qualitative issues such as saturation.  This is not correct, however, since the  problem involves multiple limits---the low density limit and the large $N_c$ limit and the heavy quark limit---and these need not commute.   Note that in the analysis done above we have first gone to the combined heavy quark and large $N_c$ limits  at  fixed $  \rho/ ( M_Q\,  \tilde{\alpha_s})^{3}  $ and then considered the  low density behavior.  The  interaction energy per baryon is then given parametrically by  $ N_c \, M_Q\,  \tilde{\alpha_s}^2$  while its density dependence is of the form $f \left (\rho / ( M_Q\,  \tilde{\alpha_s} )^{3}  \right )$ where $f$ goes to zero with $\rho$.  Suppose that there is an additional  contribution to the interaction energy baryon  which is parametrically of the form
\begin{equation}
\label{glueball}
\delta E_{\rm gb}= \Lambda_{\rm QCD} N_c^0  \tilde{\alpha_s}^2 \,  f_{\rm gb}\left (\rho \Lambda_{\rm QCD}^2 \right ) \; 
 \end{equation}
with $ f_{\rm gb}$ going to zero with $\rho$.  In the limit in which we worked, with heavy quark and large $N_c$ limits  at  fixed $  \rho/ ( M_Q\,  \tilde{\alpha_s})^{3}  $  taken prior to the low density limit, this contribution is clearly subleading.  However, if the low-density limit is taken first $\delta E_{\rm gb}$  becomes the dominant term.

If $\delta E_{\rm gb}$ is attractive at the lowest densities, the system at very low density will be unstable, ultimately collapsing to a density in which its attraction  is counteracted  by repulsive interactions $\delta E$ (at leading order in $N_c$ yielding saturated matter.   Note that as the large $N_c$ and heavy quark limits are approached, the density at which the pressure of the repulsive leading-order interaction cancels out the attraction in  $\delta E_{\rm gb}$ is pushed towards zero as is the interaction energy.     Thus, there will be a very low-density regime in which there is saturating matter, a low-density repulsive  regime where our previous calculations apply, and a 
higher-density regime where we do not presently know the equation of state of cold matter.

The preceding scenario depends on the existence of an attractive interaction energy of the form of Eq.~(\ref{glueball}) at the lowest of densities.  An equivalent way to state this is that since the mass of the baryons are heavy in our limit,  baryon kinetic energies are suppressed so that the scenario depends on the inter-baryon potential between baryons having as its longest range contribution something attractive  with a range of order $\Lambda_{\rm QCD}$.  In this combined limit the longest-range interaction is due to glueball exchange---meson exchange has a range of characteristic order of $M_Q^{-1}$ and Pauli effects at the quark level have a range $(\tilde{\alpha}_s M_Q)^{-1}$.    Note that a one-glueball exchange potential has a strength which is  parametrically of order $N_c^0$ in the $1/N_c$ expansion (which is parametrically of relative order $1/N_c$ down compared to the leading order contributions).  Moreover in the heavy quark limit, the baryon is small and thus gluonic couplings are expected to be small.   One expects that the dominant coupling of a glueball to the quarks in the baryon  will be via two gluons so that the glueball-baryon coupling will scale as $\tilde{\alpha}_s$ and the one-glueball exchange potential will scale as $\tilde{\alpha}_s^2$.  This characteristic $N_c$ and $\tilde{\alpha}_s$ dependence is seen in Eq.~(\ref{glueball}).  Now suppose  that the lightest glueball at large $N_c$ is a positive parity scalar---as we expect---the longest-range interaction is attractive and the system saturates with a saturation density  parametrically small compared to $(M_Q \tilde{\alpha}_s)^{3}$ (the natural length scale of baryon interaction).  Thus we expect that the saturation density will go  and one which goes to zero as the combined large $N_c$ and heavy quark limit is approached.

One feature might simplify the computation of the saturation density due to subleading effects: the fact that the saturation density is pushed to zero as the combined limit is approached.  This in turn means that only the longest-range  interaction due to glueball exchange---{\it i.e.}, the exchange of the lightest glueball---is relevant.

However, the problem of computing the saturation density and energy density has some non-trivial features.  Note that in the purely repulsive case studied in the previous section, the system was in a close packed crystal in order to minimize the repulsive interactions.  Once one includes the attractive non-leading but long-range effects due to glueball exchange it is not obvious {\it a priori} which crystal structure   is optimal.   Moreover, the spin-flavor effects now come into play in a more consequential way,  Note that at leading order in $1/N_c$, baryons composed of quarks of orthogonal spin-flavor states did not  interact.  However, they do feel one-glueball exchange.   There is an additional complication: in order to make a detailed calculation we need to start with a reliable energy functional.  We believe that we have a reliable energy functional for inter-baryon distances  of order $M_Q \tilde{\alpha}_s$  but that is numerically small when working in the combined limit.  We also believe that a one-glueball exchange potential is valid when working at very long distances.   However, to find the saturation density we are necessarily in a regime where the two effects yield equal and opposite pressures.   We could presumably attempt to compute this by taking as our energy functional the sum of the one from Pauli  repulsion and the one-glueball exchange.  However, from a theoretical perspective,  it is not completely  clear that such a prescription is valid for this regime.

Given these complexities we will defer any quantitative study of this problem to the future.   At this stage, we merely note the qualitative result that the system will saturate.

\section{Discussion and outlook}

In this work we have explored ``nuclear physics'' in an an artificial world in which the number of colors is large, as is the ratio of the quark masses to $\Lambda_{\rm QCD}$.   We have computed properties of both baryons and cold baryon matter in the combined limits.  While we have made significant progress in learning about this world, some important open problems remain.

The baryon's mass and form factor were computed numerically  to high accuracy in the combined limit using the mean-field formulation of Witten\cite{Witten:1979kh}. The problem at leading order is essentially solved.  However, there is an important open problem concerning corrections to the leading order result.  As discussed in Subsec. \ref{mf}  the mean-field  calculation correctly captures the leading order of the $1/N_c$ expansion (with the heavy quark limit implicitly taken at the outset).  However, it was also shown there that sub-leading terms in the $1/N_c$ expansion does not correspond to perturbation theory in the residual interaction beyond the one-body mean-field Hamiltonian.  This raises an important challenge; namely, to  formulate a tractable scheme for the systematic inclusion of  subleading terms in the $1/N_c$ expansion.  Of course, even if one can compute such corrections systematically there is still the problem of how to reliably compute corrections to the heavy quark limit.

Regarding  baryon matter,  we have derived an expression for the energy density at leading order in the combined expansion for a  phase of matter at fixed density which is taken to be on the scale of $(M_Q \tilde{\alpha}_s)^{3}$ but numerically small on this scale.  This phase is repulsive and has an exponentially small energy density.   We have shown that it is at least metastable.   We have also shown that subleading effects lead to saturating matter in which the saturation density is parametrically small on the scale of $(M_Q \tilde{\alpha}_s)^{3}$ in which the saturation density and energy density both tend to zero as the combined limit is approached.   However  there are a large number of open problems concerning baryon matter.

At a theoretical level, the most basic issue concerns the validity of the mean-field treatment.
In this work,  we assumed that a mean-field theory  treatment based on a single Slater determinant was valid in the combined limits.  This assumption was implicitly used in ref.~\cite{Witten:1979kh} in its treatment of problems with more than one baryon.  As we remarked earlier, while this assumption is  plausible, it has not been demonstrated with anything like the degree of rigor  for the single baryon problem.  Thus, the most  critical outstanding theoretical problem is to determine whether or not  this assumption is valid.   A related problem concerns corrections to the mean-field result.  As in the single baryon case, we have no tractable scheme to include $1/N_c$ corrections and corrections to the heavy quark limit.  As noted in Subsec. \ref{SL}, a viable scheme to reliably include the  longest distance effects (even though they are subleading) is essential if one wishes to compute the properties of saturating nuclear matter which arises from the interplay of  subleading but long distance attractive effects of glueball exchange with the leading order but shorter-range repulsion arising from the Pauli principle.

Apart from these theoretical issues, it is worth noting that the phenomenology of baryonic matter in this artificial world is not well understood except at low density, even in the combined limit.  Assuming that the mean-field treatment can be justified, in principle  one could  study the problem of baryonic matter numerically and thereby access higher densities.  Perhaps the most important question that needs to be addressed is whether baryonic matter saturates with densities on a scale of $(M_Q \tilde{\alpha}_s)^{3}$.   If it does, such a phase will prove to be the true ground state of baryonic matter in this regime rather than the phase arising from the interplay of  subleading, long distance attractive effects of glueball-exchange  with the leading order but shorter-range repulsion.  A final important question is whether baryonic matter undergoes any phase transitions as the density is increased.\\

This work is supported by the U. S. Department of Energy through grant number DEFG02-93ER-40762.

    \appendix
   \section{The interaction energy}
This appendix contains the derivation of Eq.~(\ref{eq:deltaE}).\\

We begin by inserting the first-order expansions of $\psi^*$, $\psi$, and $\epsilon$ in $\lambda$; Eq.~(\ref{eq:variedE}) becomes
\begin{widetext}
    \begin{equation}
\begin{split}
  &  \left[-\frac{\nabla^2}{2 M_q}+ V_0(\vec{r})-(\epsilon^0
    + \sum_i\lambda_i \epsilon_i^1)\right] (\psi^0(\vec{x})+ \sum_j \lambda_j \psi_j^1(\vec{x}))=\sum_k \lambda_k \epsilon^0 \phi^0_k(\vec{x})\\
&{\rm with} \; V_0(\vec{r}) \equiv - \tilde{\alpha}_s \int d^3\vec{r'}\;\frac{\psi^{*0}(\vec{r'})\psi^{0}(\vec{r'})}{|\vec{r}-\vec{r'}|}
\end{split}
    \end{equation}
Equating like powers of $\lambda_j$ yields:
    \begin{equation}
    \label{eq:lambdaorders}
 \begin{split}
    \lambda^0&:\;\;\left[-\frac{\nabla^2}{2 m_Q}-V_0(\vec{r})-\epsilon^0\right]\psi^0(\vec{r})=0\\
    \lambda_j^1&:\;\;\left[-\frac{\nabla^2}{2 M_Q}-V_0(\vec{r})-\epsilon^0\right]\psi_j^1(\vec{r})
    -\epsilon_j^1\psi^0(\vec{r}) =   \epsilon^0 \phi^0_j(\vec{r})
    \end{split}
    \end{equation}
\end{widetext}
    The zeroth-order equation is simply the Schr\"{o}dinger equation for the isolated baryon. It is useful to rewrite the first order equation in Dirac notation,
    \begin{equation*}
    (\hat{H}_0-\epsilon^0)|\psi_j^1\rangle-\epsilon_j^1|\psi^0\rangle=\epsilon|\phi_j^0\rangle,
    \end{equation*}
where $\hat{H}_0$ is the one-body mean-field Hamiltonian for the isolated baryon.
   Acting on both sides with the bra $\langle\psi^0|$ yields
    \begin{equation*}
-\epsilon_j^1
    =\epsilon^0 \mathcal{A}\; \; {\rm with} \; \mathcal{A} \equiv\langle  \psi^0|\phi_j^0\rangle.
    \end{equation*}
where by construction, the value of $\mathcal{A}$ is independent of $j$ since all the nearest neighbors are equivalent.

The interaction energy at leading order depends on  $\lambda_j \epsilon^1_j$.  We have already computed $\epsilon_j$.  The next step is to compute $\lambda_j$.  To proceed, we
 insert $\epsilon^1_j$  into Eq.~(\ref{eq:lambdaorders}), yielding
\begin{equation}
\left (\frac{\hat{p^2}}{2 M_q} + \hat{V_0} -\epsilon_0\right )|\psi_j^1\rangle = \epsilon_0 \left ( | \phi_j \rangle -\mathcal{A}  |\psi^0 \rangle\right ) \, .
\end{equation}
On physical grounds, we expect $\psi_j^1(\vec{r}) $ to be peaked around $r =d \hat{n}_j$ with a characteristic size of order $(\tilde{\alpha}_s M_Q)^{-1}$ since that is the type of wave function which will efficiently yield orthogonality between $\psi$ and $\phi_j$.  Let us start by doing the calculation  assuming this to be true and then verify {\it a posteriori} that the assumption was correct.  With this assumption it is easy to see  that the contribution to the left-hand side of the $\hat{V_0}$ term is characteristically of order $\tilde{\alpha}_s/d$ while the $\epsilon_0$ term is of order $\tilde{\alpha}_s^2 M_Q$.  Thus, for low densities (large $\tilde{\alpha}_s \, M_Q \, d \gg 1$) one can drop the contribution of $\hat{V_0}$  so that
\begin{equation}
\label{psi1}
|\psi_j^1\rangle = \frac{\epsilon_0}{\left( \frac{\hat{p^2}}{2 M_q} -\epsilon_0 \right )}   | \phi_j \rangle \, .
\end{equation}
where we have also dropped the term proportional to $\mathcal{A}$ since it is exponentially suppressed.  Note that the inverse operator is well defined on the space of square integrable functions since $\hat{p^2}$ is a positive operator while $\epsilon_0$ is negative and the operator is thus finite acting on any wave function.   Note, moreover, that Eq.~(\ref{psi1}) does indeed yield a wave function   peaked around $r =d \hat{n}_j$ with a characteristic size of order $(\tilde{\alpha}_s M_Q)^{-1}$; this justifies the approximation.

Recall that $\lambda_j$ enforced the orthogonality of $\psi$ and $\phi_j$:
 \begin{align*}
\left(\langle\psi^0|+\lambda_j\langle\psi_j^1|\right) \left ( |\phi_j^0\rangle + \lambda_j | \phi_j^1\rangle \right )&=0\\
    \mathcal{A}+\lambda_j\left ( \langle\psi^1|\phi_j^0\rangle +  \langle\psi^0|\phi_j^1\rangle\right )&=0\\
  \mathcal{A}+2 \lambda_j \langle\psi^1|\phi_j^0\rangle &=0\\
    \end{align*}
where the last equality follows by symmetry. We expect that the effect of the Pauli principle on each baryon in a pair of nearest neighbors to be identical.  Using this last form and our expression for $|\psi_j^1 \rangle$ yields
    \begin{equation*}
    \lambda_j=-\frac{\mathcal{A}}{2 \langle\phi_j^0|\frac{\epsilon^0}{-\frac{\hat{p^2}}{2 M_Q}-\epsilon^0}|\phi_j^0\rangle}=-\frac{\mathcal{A}}{2 \langle\psi_j^0|\frac{\epsilon^0}{-\frac{\hat{p^2}}{2 M_Q}-\epsilon^0}|\psi^0\rangle }
    \end{equation*}
  where in the last form $\phi_j^0$ in the denominator  has been replaced with $\psi^0$.  This is legitimate since the operator$(\frac{\hat{p^2}}{2 M_Q}-\epsilon^0)^{-1}$  is translationally invariant and the potential term is not present;  thus two expressions are equivalent.

The total shift in energy per baryon is thus given by:
    \begin{equation}
    \delta E \equiv\sum_j \lambda_j \epsilon_j^1=\sum_j \frac{|\langle\psi^0|\phi^0\rangle|^2}{2 \langle\psi^0|\frac{1}{\frac{\hat{p}^2}{2 M_Q}
    -\epsilon^0}|\psi^0\rangle}
    \end{equation}
which is the form of Eq.~(\ref{eq:deltaE}).

\end{document}